\documentclass[]{aa}  

\usepackage[version=4]{mhchem}
\usepackage{xcolor}

\newcommand{\transone}{NaCl $J$ = 26 $\rightarrow$ 25} 
\newcommand{\transtwo}{NaCl $J$ = 27 $\rightarrow$ 26} 
\newcommand{\magritte}{{Magritte}}
\newcommand{\vinf}{\ensuremath{\upsilon_{\text{exp}}}}
\newcommand{\Rstar}{\ensuremath{R_{\star}}}
\newcommand{\Mdot}{\ensuremath{\dot{M}}}

\newcommand{\msun}{M$_{\odot}$}
\newcommand{\Msun}{M$_{\odot}$}
\newcommand{\vlsr}{\ensuremath{\upsilon_{\text{LSR}}}}
\newcommand{\sigrms}{\ensuremath{\sigma_{\text{rms}}}}
\newcommand{\vchannel}{\ensuremath{\upsilon_{\text{channel}}}}
\newcommand{\kms}{km~s$^{-\text{1}}$}
\usepackage{multirow}
\newcommand{\ee}[1]{\ensuremath{\times10^{#1} } }
\newcommand{\err}[2]{\ensuremath{^{+#1}_{-#2}}} 
\newcommand{\Trot}{\ensuremath{T_{\text{rot}}}}
\newcommand{\edits}[1]{#1}
\newcommand{\editstwo}[1]{#1}

\usepackage{graphicx}
\usepackage{txfonts}
\usepackage[]{hyperref}
\begin{document} 

   \title{The unusual 3D distribution of NaCl around the AGB star IK Tau}


   \author{A. Coenegrachts\inst{1}
          \and
          T. Danilovich
          \inst{1,2}
          \and
          F. De Ceuster
          \inst{1}
          \and
          L. Decin
          \inst{1}
          }

   \institute{
       Department of Physics and Astronomy, Institute of Astronomy, KU Leuven, Celestijnenlaan 200D, B-3001 Leuven, Belgium
       \and
       School of Physics and Astronomy, Monash University, Wellington Road, Clayton, 3800, Victoria, Australia\\
       \email{taissa.danilovich@monash.edu}
             }

   \date{{Submitted ; accepted }} 

 
  \abstract
   {NaCl is a diatomic molecule with a large dipole moment, which allows for its detection even at relatively small abundances. It has been detected towards several evolved stars, among which is the AGB star IK Tau, around which it is distributed in several clumps that lie off-center from the star.}
   {We aim to study the three-dimensional distribution of NaCl around the AGB star IK Tau, and to obtain the abundance of NaCl relative to H$_2$ for each of the clumps.}
   {First, a new value for the maximum expansion velocity is determined. The observed ALMA channel maps are then deprojected to create a three-dimensional model of the distribution of NaCl. This model is then used as input for the radiative transfer modelling code \magritte, which is used to obtain the NaCl abundances of each of the clumps by comparing the observations with the results of the \magritte\ simulations. 
   }
   {We derive an updated value for the maximum expansion velocity of IK Tau $\vinf=28.4 \pm 1.7$ \kms. A spiral-like shape can be discerned in our three-dimensional distribution model of NaCl. This spiral lies more or less in the plane of the sky, with the distribution flatter in the line-of-sight direction than in the plane of the sky.
   We find clump abundances between $9\ee{-9}$ and $5\ee{-8}$ relative to H$_2$, with the relative abundance is typically lower for clumps closer to the star. 
   }
   {For the first time, we used deprojection to understand the three-dimensional environment of molecular emission around an AGB star and calculated the fractional abundance of NaCl in clumps surrounding the star.}


   \maketitle
%
\section{Introduction}
The asymptotic giant branch (AGB) is a late stage of the stellar evolution of stars with masses between $\sim$ 0.8 \msun\ and 8 \msun. Stars in this phase are characterized by their high mass-loss rates of $10^{-8}$~\msun~yr$^{-1}$ to $10^{-4}$~\msun~yr$^{-1}$. As a result of this mass loss, AGB stars have a circumstellar envelope (CSE) that is formed from the escaping material and which is cool and dense enough for several types of molecules and dust to efficiently form. There are three chemical types of CSEs, depending on their carbon-to-oxygen ratio: carbon-rich (C/O $>$ 1), oxygen-rich (C/O $<$ 1) and S-type (C/O $\sim$ 1).

Sodium chloride (NaCl) has been detected towards several evolved stars such as the C-rich AGB star CW Leonis \citep[][]{1987A&A...183L..10C, CW-Leo-2, CW-Leo-1}, the O-rich AGB star IK Tau \citep[][]{VY-CMa-2, Velilla-Prieto, IKTau}, and towards the red supergiants (RSG) VY Canis Majoris (VY CMa) \citep[][]{VY-CMa-2, VY-CMa-1} and NML Cyg \citep{2022AJ....164..230S}. It is a diatomic molecule with a high dipole moment \citep[$\mu = 9.0$~D,][]{DELEEUW1970288}, which facilitates its detection because it allows for efficient radiative excitation and produces bright lines. \edits{Like most metal-bearing molecules, NaCl} is highly refractory \edits{\citep{VY-CMa-1}} and it is expected to easily condense onto dust grains \citep{hofner2008,hofner2016}. 

\citet{1987A&A...183L..10C} first detected Na$^{35}$Cl (hereafter NaCl) and Na$^{37}$Cl towards an AGB star, the carbon star CW Leo.  
Based on observations from the Institut de Radio Astronomie Millimétrique (IRAM) 30m telescope, \citet{CW-Leo-2} modelled the NaCl abundance around CW Leo based on IRAM 30m observations, assuming a one-dimensional number density profile, and found a total NaCl (both $^{35}$Cl and $^{37}$Cl) abundance of $1.8\times10^{-9}$ between $3\Rstar$ and $\sim100\Rstar$ before a drop-off, suggesting that NaCl exists in both the hot inner layers and in the cooler regions of the CSE of CW Leo. They also found an isotopic ratio of \edits{Na}$^{35}$Cl/\edits{Na}$^{37}$Cl $= 2.9\pm0.3$, which is smaller than \edits{but close to} the solar ratio of \edits{$3.127\pm0.006$ \citep{Rosman1998,SolarAbundance}}.
This was confirmed by the results of \citet{CW-Leo-1}, who used the Atacama Large Millimetre/submillimetre Array (ALMA) to observe spatially resolved metal-bearing molecules in the inner regions of the CSE of CW Leo. 
They detected NaCl up to 83\Rstar\ from the star, and determined its distribution to be either a spiral or a torus. 
\citet{VY-CMa-1} used ALMA to observe spatially resolved NaCl in the CSE of the RSG VY CMa and found it to be present out to a distance of 220\Rstar, which is beyond the dust condensation radius, despite the highly refractory nature of the molecule. 
These results suggest that there is a chemical process preventing all NaCl from condensing onto dust grains, such as thermal desorption or shock-induced sputtering. Similarly to CW Leo, \citet{VY-CMa-1} detected non-spherically symmetric structures in the NaCl emission towards VY~CMa, but unlike the case of CW Leo, these take the form of separated clumps.

IK Tau is \edits{an} O-rich, Mira variable AGB star, with a moderately high mass-loss rate of $\sim 5\times10^{-6}$~\msun~yr$^{-1}$ \citep{IKTau}.
\citet{IKTau} created a spectral atlas of all species detected in an ALMA spectral scan between 335 and 362 GHz, including the relevant parameters of the detected features. Additionally, they found a complex wind morphology and both large- and small-scale inhomogeneities in the form of clumps and/or arcs which were observed in the emission of NaCl, CO, HCN, SiS and CS in the inner wind.

Thus far, NaCl towards IK Tau has been previously studied using low-spatial resolution and unresolved observations. \citet{VY-CMa-2} modelled the abundance of NaCl in the oxygen-rich CSEs of IK Tau and the RSG VY CMa using observations at 1 mm and at 2 mm of the Arizona Radio Observatory (ARO) Submillimeter Telescope (SMT) and the ARO 12m telescope respectively. They found an abundance relative to H$_{2}$ of $\sim 4\times10^{-9}$ for IK Tau, assuming a spherical distribution of NaCl, a low expansion velocity of 3 \kms, and a source size of $0.3\arcsec$.
\citet{Velilla-Prieto} observed NaCl towards IK Tau with the IRAM 30m telescope, and used population diagrams to determine its abundance relative to H$_2$, finding a value of $3.1\times10^{-7}$, around two orders of magnitude higher. They used a one-dimensional radiative transfer model and assumed local thermal equilibrium (LTE), an optically thin CSE and a source size of $0.3\arcsec$.
\citet{IKTau} detected NaCl and Na$^{37}$Cl towards IK Tau with ALMA, but not towards R Dor, an O-rich AGB star with a low mass-loss rate. The observed NaCl is not centered on the continuum peak but is observed in irregular clumps around the stellar position, while none of the other molecules they observed exhibit this behavior. 
\cite{Gobrecht} provides a theoretical prediction of the abundance of NaCl around IK Tau, by modelling the dust formation in the inner wind of IK Tau using a one dimensional, shock-induced chemical model. They predict an abundance of NaCl between $3.7 \times 10^{-10}$ and $1\times10^{-8}$ relative to H$_2$.

In this paper, we aim to study the three-dimensional distribution of NaCl around IK Tau to obtain the abundance of NaCl relative to H$_2$ for each individual clump. Sect. \ref{sec:observations} presents the observations. In Sect. \ref{sec:deprojection} we deproject the observed channel maps to obtain the three-dimensional distribution of NaCl. In Sects. \ref{sec:RTM} and \ref{sec:f}, we use the radiative transfer modelling code \magritte\ to estimate the abundance of NaCl relative to H$_2$ for each of the observed clumps. 
In Sect. \ref{sec:dis}, these results are discussed and in Sect. \ref{sec:conclusions} we give our conclusions.
   
\section{Observations}\label{sec:observations}

\begin{figure*}
    \centering
    \includegraphics[width=\textwidth]{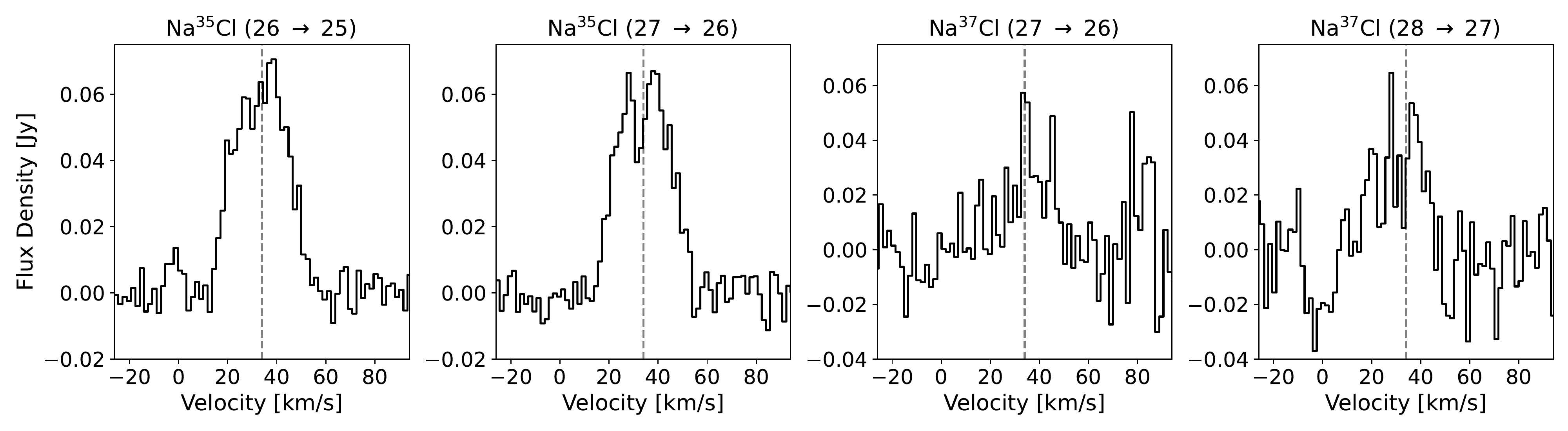}
    \caption{
        The four NaCl lines covered in the spectral scan of IK~Tau by \citet{IKTau}.
        The lines were extracted using a circular aperture with a radius of 320 mas, centred on the star. \edits{Spectra for additional extraction apertures are shown in Fig. \ref{fig:allspec}.}
    }
     \label{fig:obs_spectra}
\end{figure*}%
IK Tau was observed by ALMA in Band 7 in August 2015 (proposal 2013.1.00166.S, PI L. Decin). The measurements consist of an unbroken spectral scan between 335-362 GHz, with a spatial resolution of $\sim 120 \times 150$ mas and a frequency resolution of 1.95 MHz. The data reduction is outlined in \citet{IKTau} \edits{and was not repeated for the present work}. 
\edits{Two NaCl and two Na$^{37}$Cl emission lines were covered in the observed frequency range.}
In Table \ref{tab:transitions} we give the  quantum numbers and rest frequencies of the covered NaCl lines, along with the velocity \edits{resolution of the observations} and the \sigrms.
The lines correspond to transitions \edits{between rotational} energy levels of the ground vibrational state. 

The spectra of these lines, \edits{extracted for a circular aperture with radius 320~mas,} are shown in Fig. \ref{fig:obs_spectra}. \edits{While NaCl ($26\to25$) and ($27\to 26$) are clearly detected above the noise, Na$^{37}$Cl ($27\to 26$) and ($28\to 27$) are only tentatively detected. This is partly owing to the fact that both Na$^{37}$Cl lines lie in noisier parts of the observed spectrum \citep[see Fig. 3 in][]{IKTau}.
There were some blends with NaCl lines reported in \cite{IKTau}, which we find are not significant for the following reasons. NaCl ($27\to 26$) is reported to be blended with Si$^{18}$O ($J=9\to8$, $v=5$) at 350.959~GHz, however, this identification is tentative. With the covered Si$^{16}$O ($J=8\to7$, $v=5$) line only having a peak flux of 0.059~Jy and $^{16}$O/$^{18}$O $> 400$ for IK~Tau \citep{Danilovich2017}, the contribution from the Si$^{18}$O line to the total flux should be negligible. The only other possible blend is between Na$^{37}$Cl ($27\to 26$) and SO$_2$ ($57_{15, 43} \to 58_{14, 44}$) at 343.477~GHz, but a detailed study of the SO$_2$ emission in this dataset reports that this line was not detected towards IK~Tau \citep{Danilovich2020}.}

\edits{To more carefully examine the Na$^{37}$Cl emission, we plot all four NaCl line profiles as extracted from different regions of the channel maps. As well as the circular extraction aperture with a radius of 320~mas that we used for Fig.~\ref{fig:obs_spectra}, we also use a circular aperture with radius 800~mas, which results in noisier spectra but captures all of the NaCl flux, some of which was outside of the smaller aperture. We also extract spectra from two irregular apertures centred on the brightest two regions of NaCl flux, which we refer to as clumps A and C (see below for the complete description of clump labels). As can be seen in Fig.~\ref{fig:allspec}, clump A is characteristically blue-shifted while clump C is characteristically red-shifted. Although the Na$^{37}$Cl lines are still not clearly detected above the noise, the clump A spectra tend to show more blue emission, while the clump C spectra tend to show more red emission. This suggests that Na$^{37}$Cl behaves in broadly the same way as Na$^{35}$Cl.
We also note that the 800~mas NaCl line profiles appear to be narrower than the 320~mas profiles. This is because clump A dominates the 800~mas line profiles while being partly cropped out of the 320~mas profiles, and because the less intense redder emission is partly obscured by the noise in the 800~mas profiles. The line widths of the more extensive NaCl transitions covered by \cite{Velilla-Prieto} (with levels from $J=7$ to $J=24$) with the IRAM 30m antenna are generally reported to be narrower than other molecular lines such as CO, SiS and SiO. Although those observations are not spatially resolved, we suggest that the the narrower line profiles are a result of clump A dominating the NaCl emission generally.
}
\edits{
By comparing the integrated spectra of the $(27\to26)$ line for both Na$^{35}$Cl and Na$^{37}$Cl, we found a lower limit for the ratio Na$^{35}$Cl/Na$^{37}$Cl > 2.1, taking the different line frequencies into account as described by \cite{Danilovich2020}. This limit is comparable to the ratio found for the carbon star CW~Leo \citep{kahane2000,CW-Leo-2}. More sensitive observations would allow this ratio to be constrained with more precision.
}

\begin{table}
    \caption{Observed transitions of NaCl.}
    \label{tab:transitions}
    \centering
    \begin{tabular}{c c c c c}
    \hline\hline
    Molecule & $J^{\prime}$ $\rightarrow$ $J$ & $\nu_{\text{rest}}$ & velocity resolution & $\sigrms$\\
             &                                & (GHz)               & of channel maps     &  (mJy)\\
    \hline                        
    NaCl        & 26 $\rightarrow$ 25 & 338.0219 & $1.733$ \kms\ & 2.4\\
    NaCl        & 27 $\rightarrow$ 26 & 350.9693 & $1.669$ \kms\ & 2.6\\
    Na$^{37}$Cl & 27 $\rightarrow$ 26 & 343.4774 & $1.705$ \kms\ & 5.3\\
    Na$^{37}$Cl & 28 $\rightarrow$ 27 & 356.1436 & $1.644$ \kms\ & 6.3\\
    \hline 
    \end{tabular}
\end{table}%

\begin{figure*} 
    \centering
    \includegraphics[width = \linewidth, draft = false]{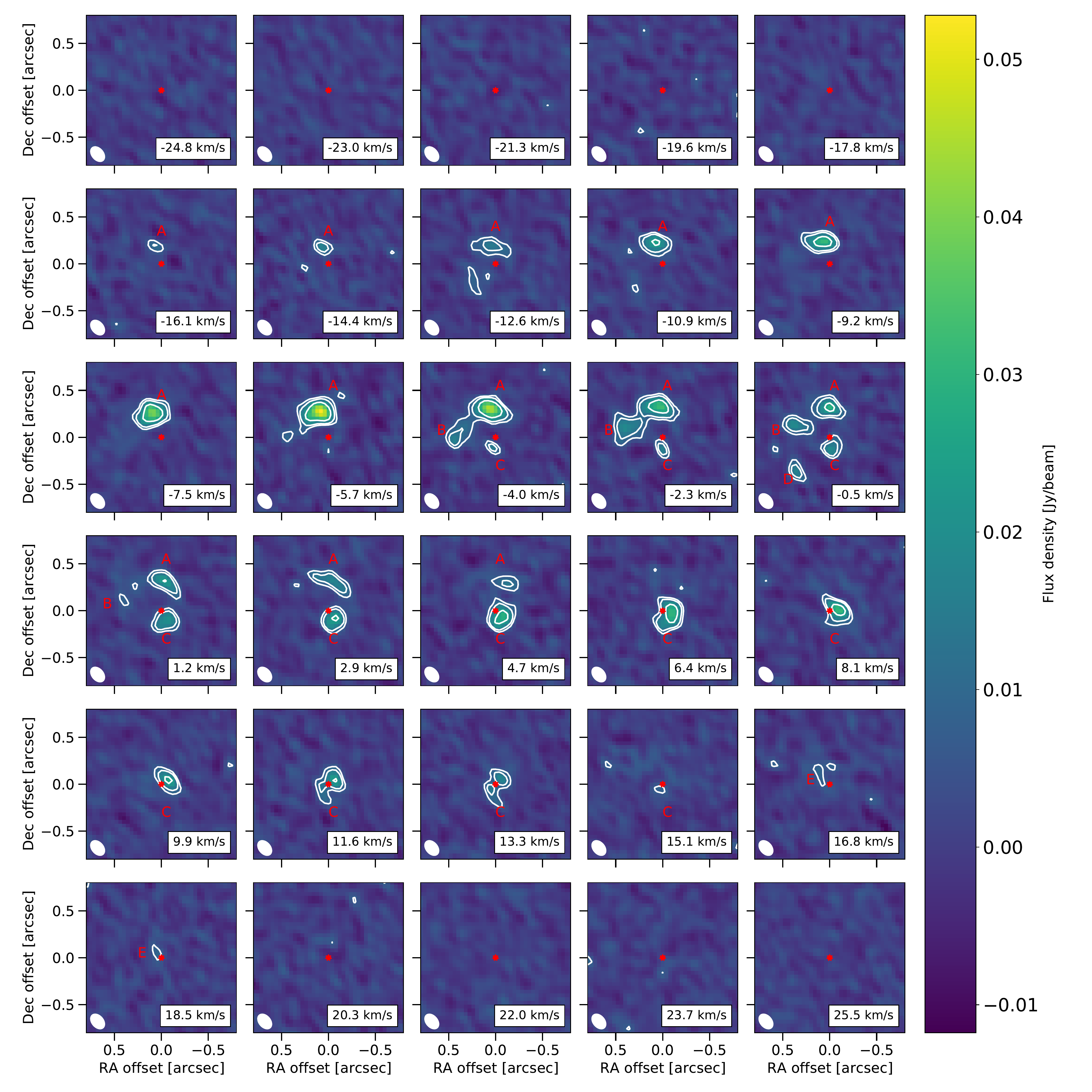}
    \caption{Channel maps of the \transone\ line. The red star indicates IK Tau's position, and the white ellipse shows the beam size. The white contours are at $3\sigrms$, $5\sigrms$, and $10\sigrms$ ($\sigrms = 2.4$ mJy). The $\vlsr = 34$ \kms{} has been subtracted from the  velocities.}
    \label{fig:channelmaps} 
\end{figure*}

Fig. \ref{fig:channelmaps} shows the channel maps of the \transone\ transition, and the channel maps of transitions \transtwo\ can be found in Fig. \ref{fig:channelmaps2}. 
The location of the star on the channel maps is indicated by a red star, and is assumed to be at the location of the continuum peak. 
The systemic velocity of the star, $\vlsr = 34$ \kms\ \citep{velprofile}, has been subtracted from the channel map velocities. 
The emission in the Na$^{37}$Cl channel maps is generally not distinguishable from the noise. Hence, we only show the spectra for these \edits{tentative lines}.

As reported by \citet{IKTau}, the NaCl emission is not centred on the star, nor is it spherically symmetric. Instead it is unevenly distributed in clumps around the star, with no clumps present to the west of the star. \edits{We identify up to five clumps comprising the NaCl emission. In Fig.~\ref{fig:channelmaps} we plot the \transone\ channel maps, and label five different clumps with the letters A to E. However,} only three of these clumps \edits{(A, B and C)} are detected in the \transtwo\ channel maps (Fig. \ref{fig:channelmaps2}). 
The lower signal-to-noise of the \transtwo\ channel maps can be explained by the \edits{slightly} higher noise at those frequencies \edits{and the slightly weaker signal (e.g. compare the spectra in Fig.~\ref{fig:allspec}, where the peak flux of \transone\ is generally 5 to 20\% brighter than the peak flux of \transtwo), possibly owing to the excitation conditions. We also note that the spectrum of clump C is the only one where \transtwo\ is 5\% brighter than \transone\ (which can also be seen by comparing Fig.s \ref{fig:channelmaps} and \ref{fig:channelmaps2}), indicating that the excitation conditions vary between clumps.}
Clumps are considered in our analysis if their $3\sigrms$ contours are larger than or roughly equal in size to the synthetic beam, or are in a channel map and spatial position directly adjacent to such a clump.
It should also be noted that clumps A and B are counted as different clumps, despite them being connected to each other, because the connection between them has a smaller cross-section than the main regions of those clumps, seeming to form a bridge between the two clumps. Additionally, clump E is only slightly larger than the synthetic beam, making it an uncertain detection, \edits{especially since it is not detected in the \transtwo\ channel maps. 
The existence of clump D is less straightforward to determine. It is only detected in one channel of \transone\ (at $-0.5$~\kms, Fig.~\ref{fig:channelmaps}) but that detection is above $5\sigma$. The $3\sigma$ contour is close to, but slightly larger than, the beam size, meaning that the clump emission spans several pixels within the $-0.5$~\kms{} channel. The non-detection of clump D in \transtwo\ could be because the velocity of the clump is very localised in the line-of-sight direction. The somewhat coarse velocity resolution of our data (1.7~\kms) could, for example, result in clump D being detected in \transone\ if it is (close to) centred on the channel velocity of $-0.5$~\kms, but not detected in \transtwo\ where its flux it might be split between the $-1.0$~\kms{} and 0.6~\kms{} channels.}

\begin{table*}
    \centering
    \caption{The minimal and maximal channel map velocities at which the NaCl clumps appear.}
    \begin{tabular}{c|c|c|c|c}
    \hline\hline
        clump    & \multicolumn{2}{c|}{\transone\ } & \multicolumn{2}{c}{\transtwo\ } \\
                 &  $\upsilon_{min}$ & $\upsilon_{max}$ & $\upsilon_{min}$ & $\upsilon_{max}$ \\
                 & (\kms)            & (\kms)           & (\kms)           & (\kms) \\
        \hline
        A        &  -16.1 & 4.7 & -14.4 & 4.0 \\
        B        &  -4.0 & 1.2 & -4.4 & 2.3 \\
        C        &  -4.0 & 15.1 & -4.4 & 14.0 \\
        D        &  -0.5 & -0.5 & \multicolumn{2}{c}{not detected above noise} \\
        E        &  16.8 & 18.5 & \multicolumn{2}{c}{not detected above noise} \\
        \hline
    \end{tabular}
    \label{tab:velclumps}
\end{table*}
 
Overall, the \transone\ channel maps have clumps detected for LSR velocities between $\upsilon = -16.1$ \kms\ and $\upsilon = 18.5$ \kms, while the \mbox{\transtwo} channel maps have clumps detected between $\upsilon = -14.4$ \kms\ and $\upsilon = 14.0$ \kms. Table \ref{tab:velclumps} shows at which velocities each labelled clump is detected. There is mostly agreement between the two transitions, although it should be noted that sometimes clumps are not detected above the noise in the \transtwo\ channel maps at velocities for which they are detected in the \transone\ channel maps, which, again, can be explained by the difference in signal-to-noise at different frequencies \edits{and the slightly weaker apparent flux of \transtwo}.
 
While the focus of the present work is on NaCl, some other molecular lines were also considered to aid our analysis. These are the SiS $J=19 \rightarrow$ 18 ($v=0$) \edits{and CS $J=7\to6$ ($v=0$) lines \citep[both analysed in detail by][]{Danilovich2019}} and the CO $J=3 \rightarrow$ 2 ($v=0$) \edits{and HCN $J=4\to3$ ($v=0$) lines}. The spectra of these lines are plotted in \ref{fig:otherspectra} and are used for the velocity determination discussed in Sect. \ref{sec:velo}.

\section{3D modelling}\label{sec:3D}
In this section, we describe how the observed channel maps can be turned into 3D models of the NaCl abundance by deprojection.
This deprojection is then used with a radiative transfer modelling code to determine the abundance of NaCl in the clumps by creating synthetic channel maps and comparing these to the observed ones.
To do this, first a new maximum expansion velocity is determined, as ALMA's increased sensitivity allows for the detection of high velocity wings that were not taken into account when the value  from the literature was determined.
Table \ref{tab:IKTau-parameters} summarises the stellar parameters of IK Tau that are used in this section.
\begin{table}
    \centering
    \caption{Stellar parameters of IK Tau. The $\upsilon_0$, $\vinf$ and $\beta$ parameters are used in the beta-type velocity law, Eq. \eqref{eq:velprofile}.}
    \begin{tabular}{llr}
        \hline\hline
        Parameter & Value & Reference\\
        \hline
         $D$ [pc] *             & $260$               & \textit{a}\\
         $T_\star$ [K]          & $2100$              & \textit{a}\\
         $\dot{M}$ [\Msun yr$^{-1}$]   & $5\times10^{-6}$    & \textit{a}\\
         $\Rstar$ [cm]          & $3.8\times10^{13}$  & \textit{a}\\
         $r_{\text{dust}}$ [cm] & $2.38\times10^{14}$ & \textit{a}\\
         $\upsilon_0$ [\kms]           & 3.0                 & \textit{b}\\
         $\vinf$ [\kms]         & $28.4 \pm 1.7^{\dagger}$    & Sect. \ref{sec:velo}\\
         $\beta$                & 1.5                 & \textit{b}\\
         $\upsilon_\mathrm{LSR}$ [\kms] & 34 & \textit{b}\\
         \multicolumn{2}{l}{Temperature profile}      & \textit{b}\\
         \hline
    \end{tabular}
    \tablefoot{References: (\textit{a}) \cite{IKTau}, (\textit{b}) \cite{velprofile}\\
    * This value lies within the errorbars of the updated value found by \cite{Andriantsaralaza_2022} based on Gaia observations. 
    $^{\dagger}$ The value of this parameter is derived in this work Sect. \ref{sec:velo}.
    }
    \label{tab:IKTau-parameters}
\end{table}%
\subsection{Expansion velocity determination}\label{sec:velo}

\edits{Prior to the higher sensitivity of ALMA observations, the terminal expansion velocity of IK~Tau was calculated to be $\vinf = 17.5$~\kms{} \citep[e.g.][from APEX and Herschel/HIFI observations]{velprofile}. \cite{IKTau} found, by fitting a velocity profile to the measured widths of various molecular lines at zero intensity, a maximal expansion velocity of $\sim25$~\kms\ for IK~Tau, owing to high-velocity wings that were not detected above the noise in earlier single-antenna observations \citep{kim, Decin2010, Decin2010a, velprofile, Velilla-Prieto}. This leaves some ambiguity as to the true terminal expansion velocity of IK~Tau\footnote{We note that IK~Tau is not unique among AGB stars in having high velocity wings. See for example the ATOMIUM sample presented in \cite{atomium}.}; the bulk of the circumstellar material appears to have a maximum expansion velocity of 17.5~\kms, and the line profiles without wings are reproduced by radiative transfer models with this expansion velocity \citep{velprofile,Danilovich2019}, but some material expands at a higher velocity. Furthermore, the underlying gas velocity field should be independent of molecular probes, i.e., although the NaCl emission is contained within 17.5~\kms\ of the LSR velocity, we should not ignore that emission from some other lines lies outside this velocity range. For example,} 
in Fig. \ref{fig:otherspectra} we plot the SiS $J=19 \rightarrow 18$ ($v=0$), CO $J=3 \rightarrow 2$ ($v=0$), \edits{CS $J=7\to6$ ($v=0$), and HCN $J=4\to3$ ($v=0$)} spectral lines, where the wings of the lines extend beyond $\vinf = 17.5$~\kms.


\edits{Our deprojection technique (see Sect. \ref{sec:deprojection}) assumes a monotonically increasing radial velocity profile. For consistency with the observed high velocity wings, we derive a new maximum expansion velocity by carefully examining the SiS $J=19 \rightarrow 18$ and CS $J=7\to6$ lines, which do not show absorption in their wings as the CO $J=3 \rightarrow 2$ and HCN $J=4\to3$ lines do (Fig. \ref{fig:otherspectra}). These lines were also chosen because their emission is thought to be approximately spherical, based on channel maps \citep{IKTau,Danilovich2019}, and because they are relatively abundant species which are not expected to have significant maser activity.}
The deprojection is done by finding the velocity corresponding to the red and blue line wings, and taking the one with the largest magnitude \citep[method outlined in Appendix A of][]{atomium}, as illustrated in Fig. \ref{fig:vinf} \edits{for the SiS line.} There, the horizontal orange dotted line at \edits{0.047 Jy/beam} corresponds to 3$\sigma$ peak rms noise. The velocities of the red and blue line wings are then determined by finding where the spectral line first goes below this criterion of 3$\sigrms$. The red wing velocity corresponds to 26.0 \kms, while the blue wing velocity corresponds to $-28.4$ \kms. The maximum expansion velocity is then the wing velocity with the greatest magnitude, which is the blue wing velocity, $\vinf = 28.4 \pm 1.7$ \mbox{\kms}. \edits{We use the velocity width of a channel map as the uncertainty. Doing the same calculation for CS, we found a lower expansion velocity $\vinf = 23.6 \pm 1.7$, probably because CS has a lower signal to noise ratio than SiS. Hence, we use the value obtained from SiS in our calculations.}

\begin{figure*}
    \centering
    \includegraphics[width = \linewidth]{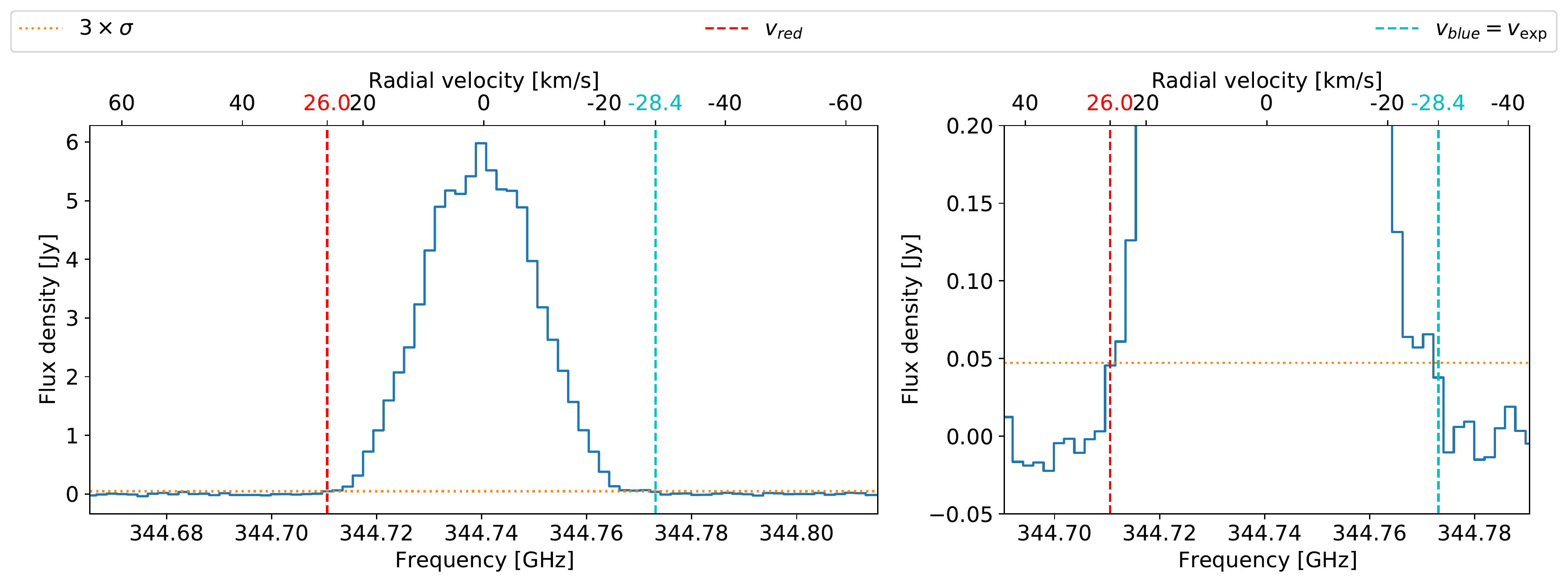}
    \caption{\textit{Left:} maximum expansion velocity determination using the SiS 19 $\rightarrow$ 18 ($v$=0). The horizontal orange dotted line is the used criterion, $3\sigrms$, the vertical dashed lines indicate the red and blue wing velocities. \textit{Right:} This is the same plot as on the left, but zoomed in to better discern the red and blue wings. Method from \cite{atomium}.}
    \label{fig:vinf} 
\end{figure*} 
\subsection{Deprojection}\label{sec:deprojection}
To model the observed NaCl clumps towards IK Tau, we deproject the channel maps, shown in Fig. \ref{fig:channelmaps}, to transform the observed (RA, Dec., $\upsilon$) coordinates to spatial ($x$, $y$, $z$) coordinates relative to the stellar position. Here, $x$ and $y$ are related to the coordinates in the plane of the sky, the right ascension and declination respectively, while the $z$ coordinate is the spatial coordinate along the line of sight. These new coordinates ($x$,$y$,$z$) all share the same units of distance, while the old coordinates have differing units (RA and Dec are angles and $\upsilon$ has velocity units). 
The spatial ($x$,$y$) coordinates of a point in a channel map can be found using the right ascension and declination offsets ($\Delta\alpha$, $\Delta\delta$) and the distance $D$ to the star (see Table \ref{tab:IKTau-parameters}):
\begin{align}
    x &= D \ \Delta\alpha \label{eq:x}\\
    y &= D \ \Delta\delta \label{eq:y}
\end{align}
Here, the right ascension and declination offsets ($\Delta\alpha$, $\Delta\delta$) are defined relative to the position of IK Tau, such that the star's coordinates on the channel maps are ($\Delta\alpha$, $\Delta\delta$) = (0,0) mas and ($x$,$y$)=($0$,$0$) AU. 
An equation for the $z$ coordinate can be calculated based on the Doppler shift of the emission, assuming a monotonic velocity field along the line of sight, see e.g. \cite{deprojection0} and \cite{deprojection}, so that:
\begin{equation} \label{eq:deproj_const}
    z = \sqrt{x^2 + y^2}\frac{\vchannel}{\sqrt{\vinf^2-\vchannel^{2}}}
\end{equation}
\edits{A full derivation of this equation can be found in Appendix \ref{sec:derivation}.} Here, $\vchannel$ is the velocity of a given channel map, and $\vinf$ is the constant, non-zero radial velocity of the stellar wind, taken to be the maximum expansion velocity. 
However, previous studies of IK Tau have used a $\beta$-type velocity law, which is a common, one-dimenional model for the accelerating gas velocity of the bulk material of the stellar wind of an AGB star \citep[see for example][]{velprofile}:
\begin{equation}\label{eq:velprofile}
    \upsilon_\beta(r) = \upsilon_0 + (\vinf - \upsilon_0)\left(1 - \frac{r_{\text{dust}}}{r}\right)^\beta
\end{equation}
Here $\upsilon_0 = 3$ \kms\ is the inner wind velocity \citep{velprofile}, generally assumed to be the sound speed; $\vinf = 28.4 \pm 1.7$ \mbox{\kms} is the maximum expansion velocity that was derived in Sect. \ref{sec:velo}; $r_{\text{dust}} = 2.38\times10^{14}$ cm is the dust condensation radius \citep{IKTau}; and $\beta = 1.5$ is a parameter determining how fast the wind accelerates \citep{velprofile}. 
The values of these parameters are also given in Table \ref{tab:IKTau-parameters}.

Since the $\beta$-type velocity law  is monotonic along the line of sight, we can replace the constant wind velocity $\vinf$ in equation \eqref{eq:deproj_const} with the velocity profile equation \eqref{eq:velprofile} 
, and get the following equation:
\begin{equation} \label{eq:deproj}
    z = \sqrt{x^2 + y^2}\frac{\vchannel}{\sqrt{\upsilon_\beta(x,y,z)^2 - \vchannel^2}}
\end{equation}
which we use to deproject the channel maps. Here, \vchannel\ is the central velocity of a channel map.
It should be noted that since the velocity profile given in Eq. \eqref{eq:velprofile} is dependent on the radial distance $r = \sqrt{x^2 + y^2 + z^2}$, the equation is implicit and needs to be solved numerically. For this, the \texttt{scipy} \citep{2020SciPy-NMeth} implementation of Brent's method \citep{brentq} was used (\texttt{scipy.optimize.brentq}).  
On a single channel map, the points with a small impact parameter $p = \sqrt{x^2+y^2}$, i.e. those that lie close to the star along the line of sight, will be deprojected to smaller $z$ values than the points with a larger impact parameter. This effect is enhanced for channel maps with velocities that are much greater than zero. 
The $z$-coordinate thus increases in size when moving away from ($x$,$y$)=($0$,$0$) AU and/or $\vchannel = 0$~\kms. \edits{This is illustrated in Fig. \ref{fig:deproj_principle}.}
 
Before the \transone\ channel maps are deprojected, a few practical adjustments need to be made. First, only the channel maps with an absolute velocity lower than the maximum expansion velocity are deprojected. No NaCl emission is detected outside of these channel maps.
Then the pixels that lie within the $3\sigma$ contours that are either larger than the size of the synthetic beam or are in a channel map and spatial position adjacent to such a $3\sigma$ contour are identified. 
The intensities of the pixels that don't lie within these contours are put to zero to avoid complicating the calculation with noise-dominated regions. After this, the coordinate transformation from ($\Delta\alpha$, $\Delta\delta$) to ($x$, $y$) (equations \eqref{eq:x} and \eqref{eq:y}) is done, and the channel maps were cut down to a smaller size in $x$, $y$, encompassing all the clumps.
 
Next, the intensity is interpolated to a grid that is 6 times as dense in the $x$ and $y$ directions, and 12 times as dense in the $\upsilon$ direction, yielding a grid of $132 \times 156 \times 300$ points. 
The higher grid-density in the $\upsilon$ coordinate is chosen because there are large spacings in the $z$ direction of the ($x$, $y$, $z$) grid corresponding to the given ($x$, $y$, $\upsilon$) grid.
As mentioned before, the $z$ coordinate will have a larger (smaller) absolute value when the $x$, $y$, and/or $\upsilon$ coordinates have absolute values far from (close to) 0. This leads to a non-constant spacing in the $z$ direction, which can be large for grid points far from the origin. The higher grid-density reduces the distance between grid points in the $z$-direction.
This grid is then deprojected using Eq. \eqref{eq:deproj}. It is later used as an input for the radiative transfer modelling code \magritte\ (see Sect. \ref{sec:RTM}), and the higher grid density allows for more granular mapping of quantities such as molecular number density and temperature. 
 
\begin{figure*}
    \centering
    \includegraphics[width = \linewidth]{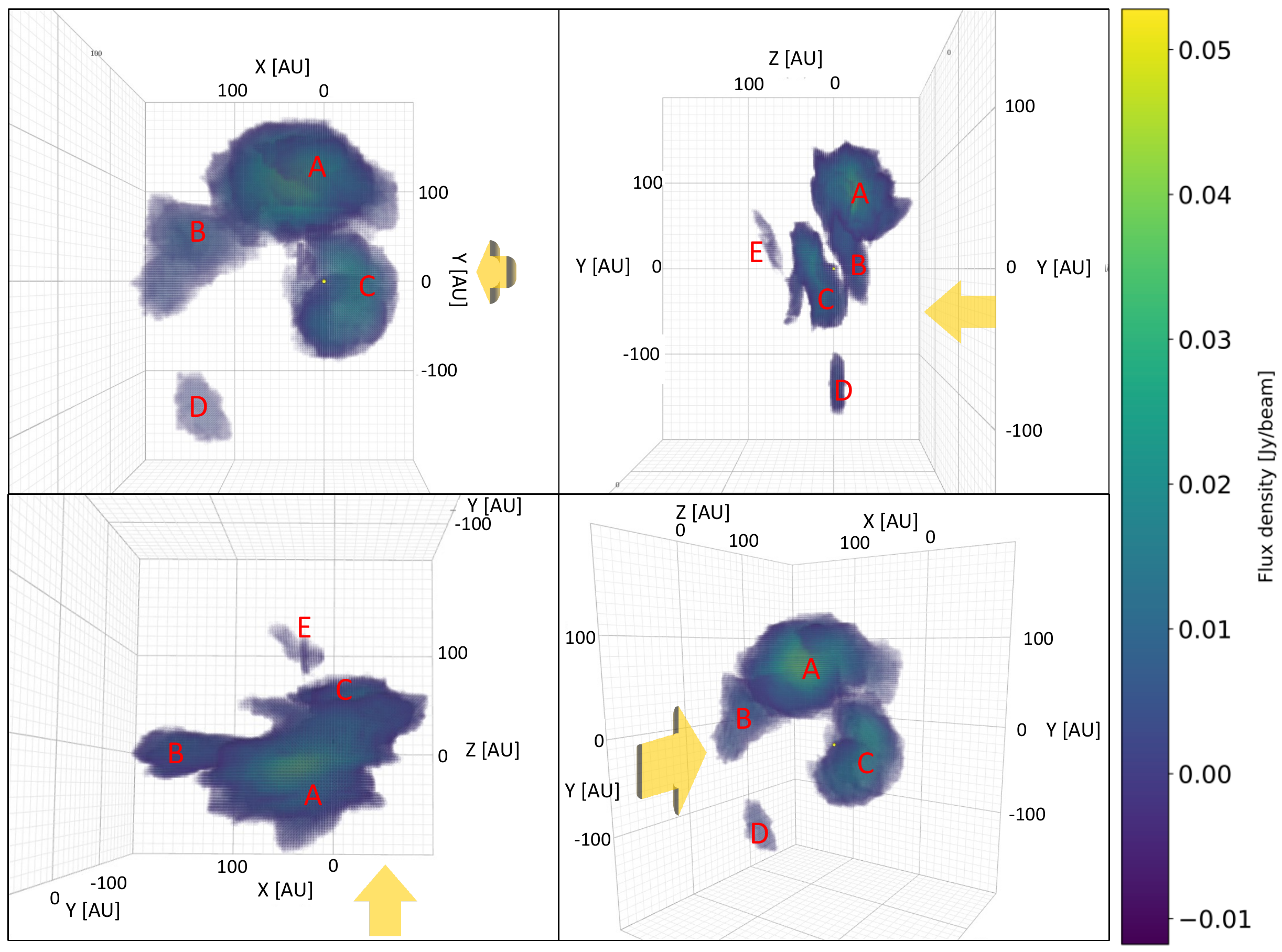}
    \caption{Four panels showing the deprojection of the \transone\ channel maps from different angles. Only points detected at $\geq3\sigma$ are shown, and the colors indicate intensity, using the same colorbar as for the original channel maps (Fig. \ref{fig:channelmaps}). The yellow arrow points along the direction of the line of sight. 
    }
    \label{fig:deproj} 
\end{figure*}

Fig. \ref{fig:deproj} shows the deprojected clumps. The colors relate to the observed intensity at those points, and the little yellow dot shows the central star, IK Tau, to scale \citep{IKTau}.
The observed structures extend about $\sim$ 200 AU in the $x$-direction, $\sim$ 240 AU in the $y$ direction and $\sim$ 140 AU in the $z$-direction, meaning that the structures are flatter in the $z$-direction, given the assumed $\upsilon_\beta$ law.

\subsection{Radiative transfer modelling}\label{sec:RTM}
The channel maps (Figs. \ref{fig:channelmaps} and \ref{fig:channelmaps2}) and our deprojection show that the distribution of NaCl is very asymmetric, so a one-dimensional radiative transfer model cannot be used to study the emission. Instead, the three-dimensional radiative transfer code \magritte\footnote{ \url{https://github.com/Magritte-code/Magritte} } \citep{magritte_I,magritte_II} is used.
\magritte\ computes the radiation field by solving the radiative transfer equation along rays (straight lines) through the medium. For numerical stability, \magritte\ solves the radiative transfer equation in a second-order, Schuster-Feaurier form \citep{schuster, feautrier}. For more technical details, see \cite{magritte_I}. 
 
The input for \magritte\ consists of 1) a three-dimensional point cloud with a) spatial coordinates, b) a velocity field, c) a temperature profile,  d) the number densities of all considered molecular species, e) a turbulent velocity field, 2) the boundary of the grid, 3) the transition rates of the considered molecular species
, and 4) the frequencies at which it needs to compute the output.
The deprojected grid described in Sect \ref{sec:deprojection} is used as the three-dimensional grid with spatial coordinates, after it has been interpolated a second time to a new grid of $132 \times 156 \times 115$ points. This grid gives the $z$ direction the same spatial scale as the $x$ and $y$ directions, such that the grid is evenly spaced, and crops points representing empty space to make the model more efficient. To reduce computation times further, the \magritte\ model is remeshed \citep{magritte_II}, meaning that the grid will be resampled to have more points in the high density regions and have fewer points in the low density regions. 
\edits{The reduced \magritte\ model has a spherical inner boundary centred on the star, with a radius equal to the stellar radius ($\sim 2.54$ AU), and a cuboidal outer boundary.}

The radiative transfer code requires the three vector components of the velocity profile rather than the total radial velocity (calculated using Eq. \eqref{eq:velprofile}). Assuming a radially outward spherically symmetric outflow, the components are given by:
\begin{equation}
    \upsilon_x = \upsilon \ \frac{x}{r} ,
    \qquad 
    \upsilon_y = \upsilon \ \frac{y}{r} ,
    \qquad 
    \upsilon_z = \upsilon \ \frac{z}{r} ,
\end{equation}
with $\upsilon_i$ the component of the velocity along the $x$, $y$ or $z$ direction, $\upsilon$ the radial velocity at point ($x$,$y$,$z$), and $r = \sqrt{x^2+y^2+z^2}$.  

\begin{figure}
    \centering
    \includegraphics[width = \linewidth]{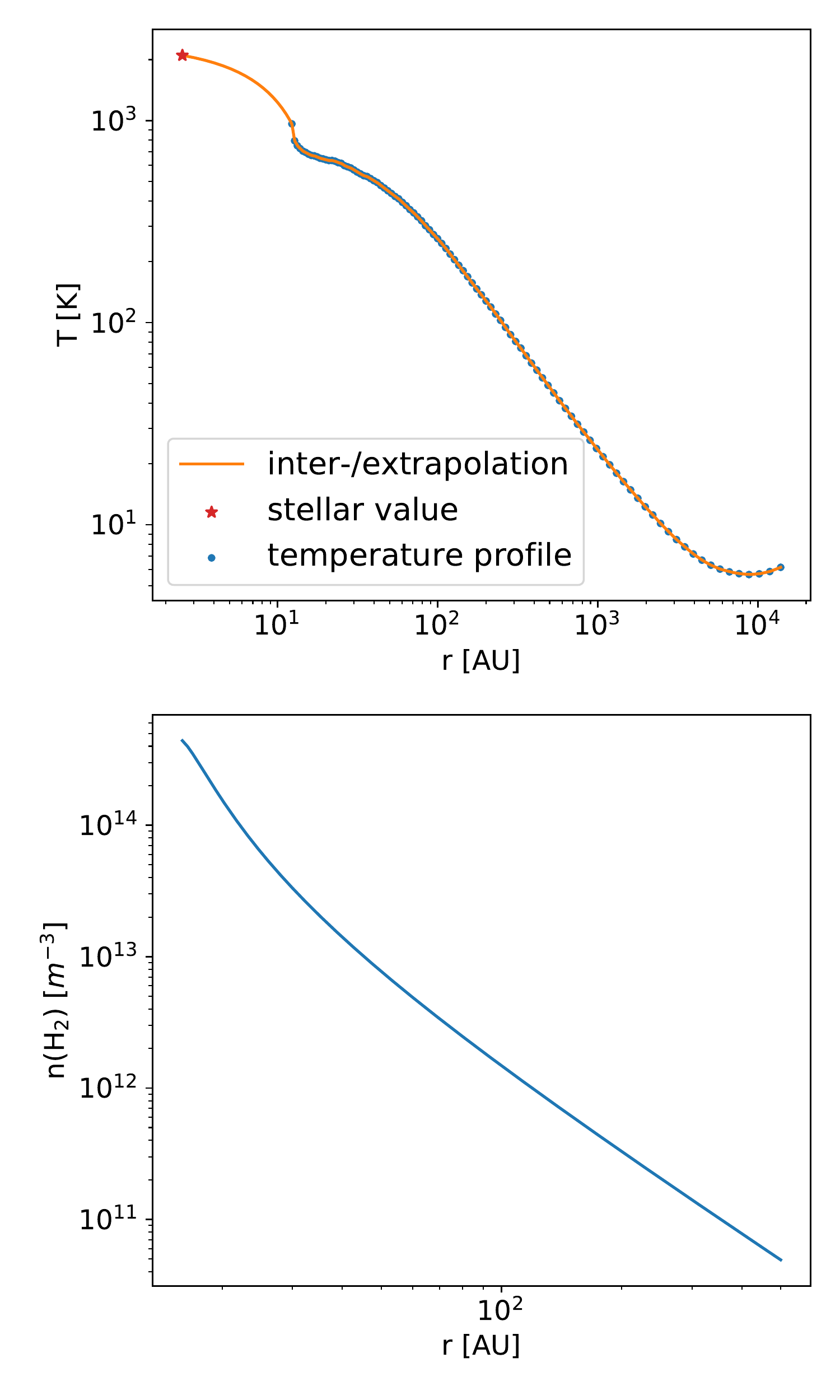}
    \caption{\textit{Top:} the temperature profile derived by \cite{velprofile}, extrapolated to the stellar radius and effective temperature of IK Tau. \textit{Bottom:} the number density profile of H$_2$ calculated from the mass-loss rate using equation \eqref{eq:nH2}.}
    \label{fig:nT} 
\end{figure}

We use a temperature profile based on the results of \cite{velprofile}. The inner boundary of this temperature profile is the dust condensation radius $R = 18\times10^{13}$ cm, which is not small enough for our purposes. Therefore, we linearly extrapolate it to match the effective temperature of the star at $r = \Rstar$. The extrapolated temperature profile can be seen in the top panel of Fig. \ref{fig:nT}. 
 
In the model, we consider H$_2$ and NaCl, where H$_2$ serves as an approximation of the total circumstellar gas. Since the H$_2$ abundance is typically much greater than any other molecular species, its number density can be approximated from the mass-loss rate of the star as follows, assuming a spherically symmetric outflow and that hydrogen is mainly in its molecular form H$_2$:
\begin{equation}\label{eq:nH2}
    n(\text{H}_2) = \frac{\rho(\text{H}_2)}{m(\text{H}_2)} = \frac{1}{m(\text{H}_2)}\frac{\Mdot}{4\pi r^2 \upsilon_\beta(r)}
\end{equation}
where $\upsilon_\beta(r)$ is the gas expansion rate at radial distance $r$, which is calculated using the beta-type velocity law Eq. \eqref{eq:velprofile}.
The radial dependence of this formula is plotted in the bottom panel of Fig. \ref{fig:nT}.

We assume Local Thermodynamic Equilibrium (LTE) in our radiative transfer modelling and we include only the ground vibrational levels. The modelling time would increase by about four orders of magnitude without the assumption of LTE, with an additional multiplier for higher vibrational states. We expect that models using non-LTE would provide more accurate results, especially further from the star, where the density is low. 
Infrared radiative pumping might have a significant contribution, but this cannot be taken into account for the model without including more vibrationally excited levels.
The radiative transition rates and energy levels for NaCl are taken from the ExoMol line list database \citep{exomol}, and the collisional rates for NaCl were calculated using the formulation provided by \cite{CW-Leo-1}. We consider 40 radiative transitions between 41 energy levels of NaCl.

By construction, the observations are binned in frequency channels yielding one channel map per frequency bin.
As a result, any variation of the intensity within a frequency bin is integrated out, and we only observe the integrated intensity over the frequency bin.
The radiative transfer solver, however, constructs the synthetic channel maps by computing the intensity at specific frequencies.
Therefore, any variation of the intensity in between those frequencies is not necessarily integrated out, and we obtain the intensity at a certain frequency rather than the integrated one over a frequency bin.
To simulate the integration over the frequency bins, we computed intensity maps for three equally-spaced frequencies within each bin, and summed the intensity maps.
An example of these channel maps can be found in Fig. \ref{fig:synmaps1} in appendix \ref{sec:A}.


The resulting synthetic channel maps don't yet take the ALMA beam size into account. This can be resolved by using the Common Astronomy Software Applications for Radio Astronomy \citep[CASA,][]{CASA}, which is a data processing software for radio interferometers, including ALMA. By using the \texttt{simalma} tool, it is possible to simulate what the synthetic channel maps from \magritte\ would look like if they were actually observed by ALMA. IK Tau was observed when the percipitable water vapour (pwv) was in the range [0.2, 1.7] mm. To account for this, we run \texttt{simalma} for both values at the edges of the range. An example of these channel maps can be found in Fig. \ref{fig:synmaps2} in Appendix \ref{sec:A}.

Due to the trimming that was described earlier in this section, the angular extent of the resulting synthetic channel maps is too small for CASA to run efficiently. To solve this, a border which has the exact same value as the background is added, extending the channel maps.

\subsection{NaCl abundance determination}\label{sec:f}
The number density of NaCl within each clump is defined relative to H$_{2}$, i.e. $n(\text{NaCl}) = f_i\times n(\text{H}_{2})$ with $f_i$ the NaCl abundance of clump $i$. 
As a first approximation, we used the NaCl abundances from the literature, which are listed in Table \ref{tab:NaCl_lit}.
Within a clump, the number density is assumed to be constant, and we define $n(\text{H}_{2})$ at the centre, $r_{\text{centre}}$, of each clump, using equation \eqref{eq:nH2}. For each clump, Table \ref{tab:clump_dist} lists the distance to IK Tau from both the centre of the clump $r_{\text{centre}}$ (calculated as the median of the distances of all the points making up the clump) and the closest point to the star $r_{\text{closest}}$.
At the grid points outside of the clumps, the NaCl number density has an arbitrarily small value of $10^{-40}\times n(\text{H}_{2})$, as \magritte\ requires a number density greater than zero.

\begin{figure}
    \centering
    \includegraphics[width = \linewidth, draft = false]{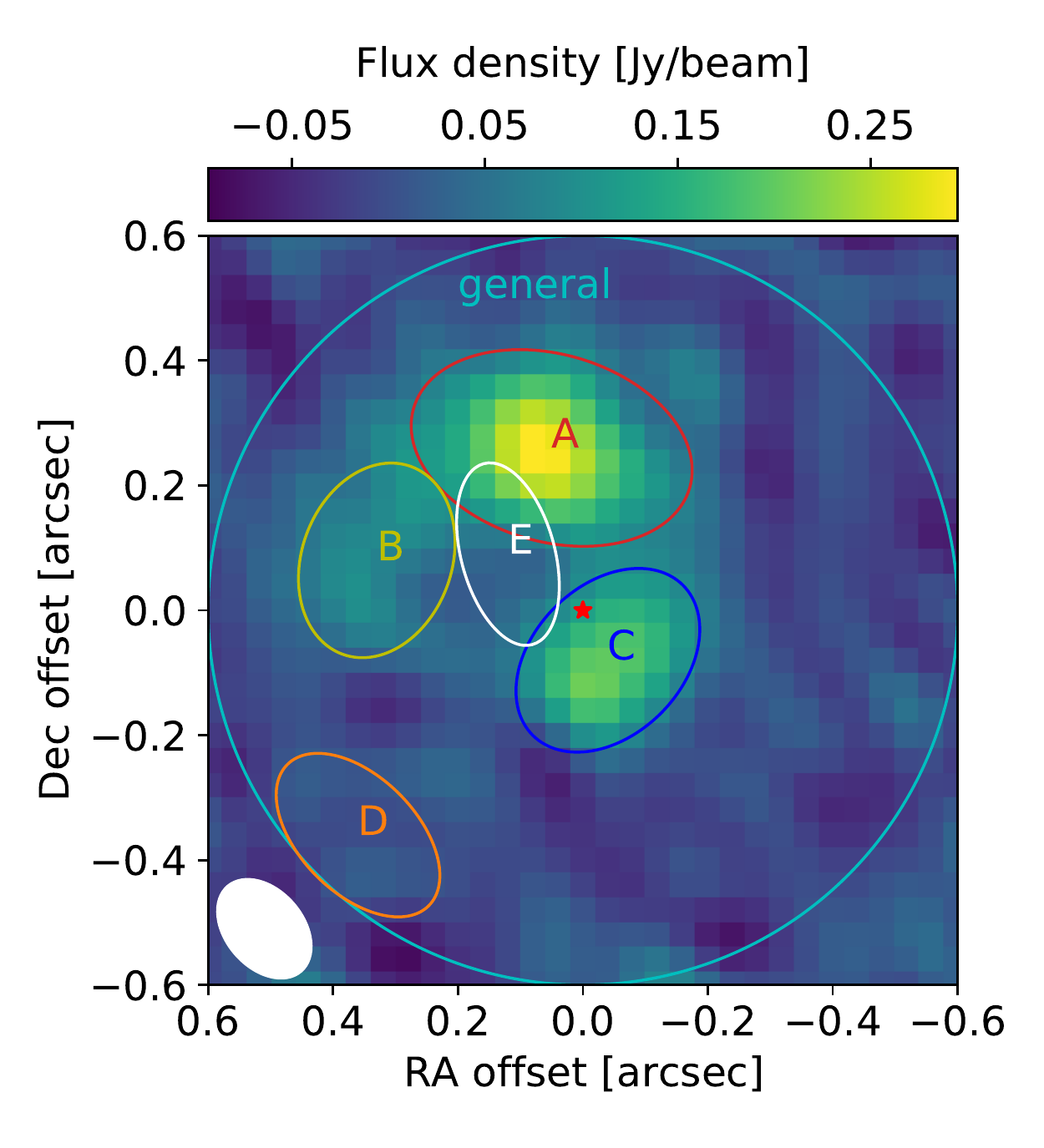}
    \caption{The apertures used for extracting spectra from the observed and synthetic channel maps. The ellipses labelled A through E represent each of the clumps, and the big circle is the general emission area. Areas A, B, C and general are also used for the rotational temperature diagram (see Sect. \ref{sec:RTD}). The background is the zeroth moment map of the \transone\ transition, where yellow (blue) is the highest (lowest) intensity, and the white ellipse is the beam.
    }
    \label{fig:areas} 
\end{figure}

\begin{table}
    \centering
    \caption{The relative abundances of NaCl relative to H$_2$ towards IK Tau from three different literature sources.}
    \begin{tabular}{cc}
    \hline\hline
        Source & $f$ \\
        \hline
        \cite{VY-CMa-2} & $4\times10^{-9}$ \\
        \cite{Gobrecht} & $3.7\times10^{-10} - 1\times10^{-8}$ \\
        \cite{Velilla-Prieto} & $3.1\times10^{-7}$ \\
        \hline
    \end{tabular}
    \label{tab:NaCl_lit}
\end{table}
\begin{table}
    \centering
    \caption{The distance from IK Tau to each of the clumps.}
    \begin{tabular}{c|ccccc}
    \hline\hline
    Clump \#           & A & B & C & D & E\\
    \hline
    $r_{\text{centre}}$  [AU] & 84.6 & 106.7 & 41.0 & 133.4 & 68.4 \\
    $r_{\text{closest}}$ [AU] & 38.6 & 51.4 & 3.0 & 114.2 & 52.0 \\
    \hline
    \end{tabular}
    \label{tab:clump_dist}
\end{table}%
A grid of \magritte\ simulations was run for different clump abundances. For each model, the intensities of the resulting synthetic channel maps were compared to the observed ones. 
To make an unambiguous comparison, spectra were extracted for regions covering each of the clumps, and for a region containing all the clumps of both the observed and synthetic channel maps. These regions are plotted in Fig. \ref{fig:areas}. The following $\chi^2$ statistic was then calculated to find the best model.
\begin{equation} \label{eq:chi2}
    \chi^2= \sum^N\frac{(I_{\text{spectrum}} - I_{\text{obs}})^2}{\sigma^2}
\end{equation}
Here, $I$ is the integrated line intensity, $N$ is the number of spectra ($N=6$ for five clumps and one total region), and $\sigma$ is the uncertainty on the observed spectra, which is $\sigma = 7\%$ \citep{IKTau}.

\begin{figure}
    \centering
    \includegraphics[width = \linewidth, draft = false]{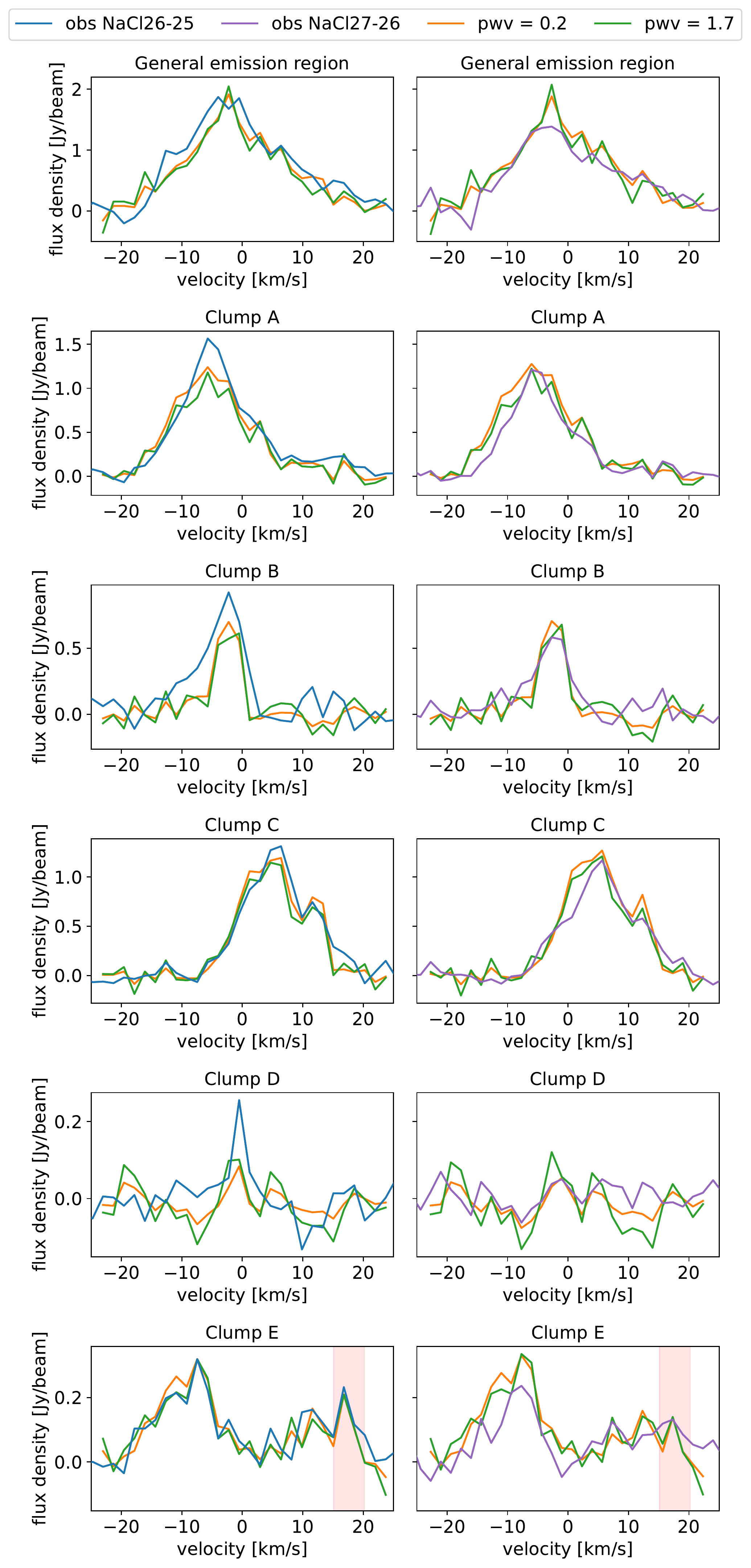}
    \caption{
    Synthetic spectra of transitions \transone\ (left) and \transtwo\ (right), extracted from the emission region of each of the clumps, and of the general emission region. Fig. \ref{fig:areas} shows the used extraction apertures. For clump E, the only relevant peak is indicated in red, as the rest is caused by the overlap with the emission region of other clumps.
    }
    \label{fig:synspec2} 
\end{figure}

We model transitions \transone\ and \mbox{\transtwo} at the same time, calculating the $\chi^2$-function across all clumps and transitions.
To explore the parameter space, we first set the abundance of all clumps to $f = 3.1\ee{-7}$ \citep{Velilla-Prieto}, which we find to be too large. Then, we set the abundance to $f = 4\ee{-9}$ \citep{VY-CMa-2}, which we find is too small. Thereafter, to determine the best fitting abundance for each clump, we varied the abundances between these values. 
Fig. \ref{fig:synspec2} shows the spectra of the final model for both lines, and Table \ref{tab:f2} gives the retrieved abundances and the corresponding number densities. The uncertainties in this table are for a confidence interval of 50\%. 
\begin{table*}[]
    \centering
    \caption{The retrieved clump abundances considering both \transone\ and \transtwo. The errors are given for a confidence interval of 50\%.}
    \begin{tabular}{c|c|ccccc}
    \hline\hline
        &pwv & clump A & clump B & clump C & clump D & clump E \\
        & (mm) \\
        \hline 
        \multirow{2}{*}{$f$ [$10^{-8}$]}         & 0.2 & $4.4\err{0.6}{0.7}$ & $4.6\err{0.7}{0.8}$ & $1.0\pm 0.2$ & $4.0\err{0.5}{0.4}$ & $2.8\err{0.6}{0.5}$ \\[1ex]
         & 1.7 & $3.8\err{0.7}{0.6}$ & $4.1\err{0.9}{0.6}$ & $0.9\pm 0.2$ & $4.9\err{0.5}{0.4}$ & $2.6\err{0.6}{0.5}$ \\[1ex]
        \hline
        \multirow{2}{*}{$n$ [$10^{4}$ m$^{-3}$]} & 0.2 & $9.5\err{1.3}{1.5}$ & $5.9\err{0.9}{1.0}$ & $13 \pm 3  $ & $3.1\err{0.4}{0.3}$ & $10.0\err{2.1}{1.8}$ \\[1ex]
         & 1.7 & $8.2\err{1.5}{1.3}$ & $5.3\err{1.2}{0.8}$ & $12 \pm 3  $ & $3.8\err{0.4}{0.3}$ & $9.3\err{2.1}{1.8}$ \\
    \hline
    \end{tabular}      
    \label{tab:f2}
\end{table*}


\section{Discussion}\label{sec:dis}
\subsection{Deprojection}\label{sec:dis-deproj}
\subsubsection{Choice of velocity profile}

\edits{In this work, we have assumed, as have previous studies \citep{Decin2010,velprofile,IKTau}, a beta-law expansion velocity profile (equation \ref{eq:velprofile}) for the CSE of IK~Tau. This ensures a monotonically increasing radial velocity, yielding an invertible projected velocity along the line of sight, which is necessary for our deprojection method. However, it is possible that the CSE of IK~Tau has a more complex velocity field. For example, there may be shocks propagating through the CSE --- especially in the inner regions close to the star where pulsations could drive shocks --- or the wind may be partially shaped by a stellar or planetary companion, affecting the velocity field gravitationally.
In principle, one could model these effects and as long as the resulting velocity along the line of sight  is invertible, one could still employ our deprojection method. Furthermore, from the synthetic observations of the deprojected model, one could then fit the additional model parameters.
However, this would greatly extend the parameter space, and it would be very difficult to impose the invertibility of the velocity along the line of sight.
In future work, we will explore these possibilities with another (probabilistic) deprojection method that relies on fewer and less strict assumptions.
}

\edits{We tested a few different values of \vinf\ in our deprojection, including $\vinf = 17.5$~\kms, the value derived from single-dish observations, and values between this and our final value of $\vinf = 28.4$~\kms. In general, assuming larger values of \vinf\ in equation \ref{eq:deproj} results in a more compressed 3D distribution in the $z$-direction. If \vinf\ is smaller than the largest absolute channel velocity considered, equation \ref{eq:deproj} will have an imaginary denominator and the deprojection will fail. The physical implication of this is that if the high velocity wings are created by processes (such as shocks) close to the star, which we consider to be more likely than the wings resulting from some gas being accelerated at the edge of the wind, then the emission corresponding to the high velocity wings will be placed on an outer edge in the deprojected model and will never be placed close to the star (in the $z$-direction). This has some consequences for the 3D structure of the NaCl distribution and, in particular, for the location of clump E, which we discuss in more detail in the following section.
}

\edits{Finally, we reiterate that whatever the true velocity field may be, it should be equally applicable to all the molecular emission in the CSE of IK~Tau, although different molecular lines may trace different parts of the velocity field. The main driver of the CSE expansion is radiation pressure on the dust which then drags the gas with it \citep{hofner2018}. With this in mind, there is no apparent reason why NaCl should have different kinematic behaviour to other molecules. The asymmetric distribution points to an unusual formation mechanism, but the distribution of the clumps suggests that they are expanding with the rest of the CSE, especially if we consider clump C to have formed most recently, with clumps A, B, and D getting progressively older (see more detailed discussion on clump ages below).
}

\subsubsection{3D structure of NaCl}\label{sec:3D-struct}
In the three-dimensional model of the NaCl distribution around IK Tau (Fig. \ref{fig:deproj}), a tentative spiral-like shape can be discerned from the clumps, starting from clump C, moving outwards through clumps A and B to end in clump D. This spiral-like shape lies more or less in the $xy$-plane 
and could be caused by a companion, although none has been directly observed as of yet. The structures seen in the ALMA CO data of IK Tau are also indicative of a spiral-like structure \citep{IKTau}. Another possible cause of the spiral-like pattern could be rotation of the star, \editstwo{much like the sprinkler-like scenario described by \cite{Quintana-Lacaci2022}. Other possible scenarios for the formation of the observed distribution of NaCl are discussed in Sect.~\ref{sec:chem}}. 

\editstwo{Although we have generally treated the clumps of NaCl as discrete in this work, their apparent separation may be an observational limitation. That is, lower abundances (corresponding to lower intensities) of NaCl may be present between the clumps we have defined here. In Fig.~\ref{fig:deproj} it can already be seen that clumps C and A, and A and B are almost touching. A less abundant trail of NaCl may in fact be linking them (and also clumps B and D) while not being detected above the noise of our observations. The clumps we do detect may hence be enriched regions of NaCl along a putative spiral.}

\begin{figure*}
    \centering
    \includegraphics[width = \linewidth, draft = false]{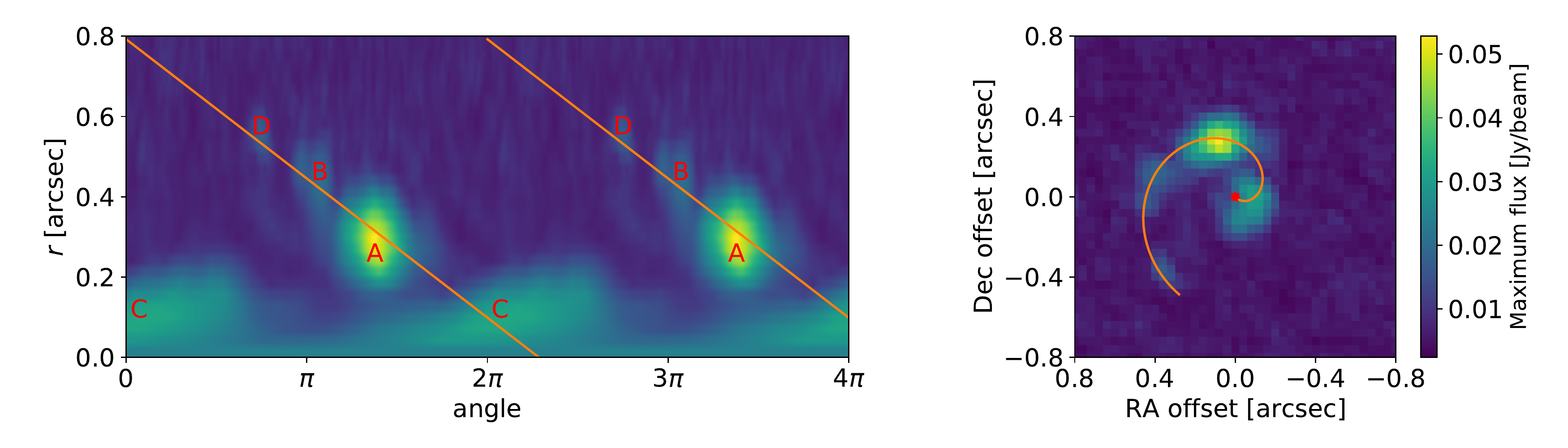}
    \caption{\edits{On the left we plot the radial emission distribution against the angle for the maximum intensity plot, with a full revolution repeated three times. The clumps are labelled and an orange line is fit to the brightest point in each clump. On the right we show the maximum intensity plot with the same line, now appearing as a spiral passing through clumps C, A, B and D.}
    }
    \label{fig:spiral}
\end{figure*}

\edits{To better understand the \editstwo{putative} spiral-like structure, we examine the maximum intensity plot of \transone.
The maximum intensity plot was made \cite[using the CASA software,][]{CASA} by selecting the maximum flux for each pixel in any channel across the NaCl line (i.e. the range of channels in Fig.~\ref{fig:channelmaps}). In this way, the spatial locations of all four main clumps (A, B, C, and D) can be seen in a single plot, whereas clumps B and D are difficult to discern in an integrated intensity map such as the zeroth moment map shown in Fig.~\ref{fig:areas}.
In Fig.~\ref{fig:spiral} we plot the radially distributed emission of the maximum intensity map as a function of angle for a circle centred on the star. We show the plot for three full revolutions and label the four visible clumps (A to D). We fit a straight line to the brightest points in each clump and then plot this fit over the maximum intensity plot in the usual RA and Dec coordinates, with the result of a spiral fit to the clumps. The fact that clump D lies close to the LSR velocity (see Fig.~\ref{fig:channelmaps} and discussion in Sect.~\ref{sec:3D-struct}) indicates that the spiral arrangement of the clumps lies close to the plane of the sky, though the offset velocities of clumps A and C in opposite directions suggest that the spiral might be tilted along the axis between clump D and the central star, such that clump A is angled towards us and clump C is angled away from us by $31\pm11^\circ$ (with the uncertainty based on the beam size).
}

The relation between the size of a clump and the distance to the star can tell us something about the evolution of the clumps in time and how the NaCl abundance varies between them.
The distances to each of the clumps are given in Table \ref{tab:clump_dist}. 
Clump D is one of the smallest clumps and the farthest from the star, while clump C is both the largest and the closest to the star. Similarly, clump A is slightly closer to the star than clump B, and is larger as well. Given that the CSE in which these clumps lie is radially expanding, it can be assumed that the closer a clump is located to the star, the more recently it was formed, and vice versa. This implies that clump D is the oldest, and clump C is the youngest. 
Two possible explanations for the decreasing size with distance can be invoked: (1) the clumps expand as they travel outwards, making the outer regions of the clumps too diluted to be detected, or (2) NaCl is destroyed more readily farther from the star, either through condensation onto dust or via photodissociation.
In both cases, the NaCl number density of a clump would decrease as it travels outwards. 
We thus expect clump D to have the lowest NaCl number density and abundance, while clump C would have the highest. We find that this holds for the NaCl number density, but not for the fractional abundance: the clumps further from the star, such as clump D, have a higher abundance than the clumps closer to the star, such as clump C. This is discussed in more detail in Sect. \ref{sec:dis-f}.

The one outlier is clump E, which is similar in size to clump D, but which lies at a similar distance to clumps C and A. 
Two possible explanations for this are: (1) because the contour defining clump E is only slightly bigger than the beam size (see Sect. \ref{sec:observations}), it could possibly be misinterpreted noise; (2) clump E could also be a part of clump C, similarly to how clumps A and B are connected. In the top right panel of Fig. \ref{fig:deproj}, it can be seen that clump C already has a similar protrusion in the negative $y$-direction, so clump E could be connected to clump C in the same way.
\edits{To expand on this idea, it is possible that clump E and the negative-$y$ protrusion of clump C are not located where our deprojection model (with the monotonic radial velocity profile) places them, but are instead high velocity components located close to the star. The observations presented here were taken close to the peak of IK~Tau's pulsation phase \citep{Decin2017}. It could be that these two features are part of a shock that has accelerated the gas in a relatively small region close to the star. (Such a shock could also account for the high velocity wings seen in other molecular line profiles, but a detailed analysis of other molecular emission is beyond the scope of the present work.)}


\subsection{Origin of the NaCl clumps}

\edits{No other molecules detected towards IK~Tau with ALMA show the same kind of irregular emission as NaCl. If one of the more abundant species were to display such clumpy emission, it is possible that the clumps could be obscured by the rest of the more extended and symmetric emission. However, an examination of the data, particularly of CO, SiO and HCN, does not reveal any evidence of clumps or overdensities corresponding to the NaCl clumps. Other metal-bearing molecules that have been (tentatively) detected towards IK~Tau, such as TiO \citep{Danilovich2020a}, AlCl and AlOH \citep{Decin2017}, exhibit emission spatially centred on the star. This includes TiO, which has a blue-shifted spectral line profile, similar to what we see for NaCl owing to the dominance of clump A, but spatially centred emission \citep{Danilovich2020a}, unlike NaCl.
}

Our results can be compared to the carbon-rich AGB star CW~Leo \citep{CW-Leo-1} and the red supergiant VY~CMa \citep{VY-CMa-1,Quintana-Lacaci2022}, which both exhibit structures in their NaCl emission. \cite{CW-Leo-1} found that the emission of NaCl and KCl around CW~Leo could be explained by either a spiral or a torus. However, this emission is \edits{predominantly} centred on the star \edits{and most of the emission is part of a continuous structure in each channel}. This is not the case for IK~Tau \edits{and could suggest a different formation mechanism in IK~Tau than for CW~Leo}. 
For VY~CMa, \cite{VY-CMa-1} found multiple clumps, whose total emission is not centred on the star, with two clumps lying at least 1\arcsec{} farther from the star. They suggest that localised mass ejections that drive sputtering off dust could be the origin of these clumps. \edits{\cite{Quintana-Lacaci2022} obtained higher resolution observations of molecular emission towards VY~CMa and found that the emission from NaCl and H$_2$S trace jets with a Hubble-like velocity field. We see no evidence of jets around IK~Tau. Although there are no spatially resolved observations of H$_2$S towards IK~Tau, a previous study based on APEX data found much wider H$_2$S line profiles than those of NaCl and modelling indicated a relatively large emitting envelope centred on the star \citep{Danilovich2017a}. This is a strong indication that NaCl and H$_2$S do not trace the same regions around IK~Tau and that one or both likely do not have the same formation mechanisms around IK~Tau as around VY~CMa.}

\edits{For IK~Tau, the irregular distribution of the NaCl clumps and their apparent spiral distribution suggests a temporally variable process is behind their formation. We considered several possible contributing scenarios for the formation of the NaCl clumps, which we outline in the following subsections. These include the impact of pulsations, correlations with the continuum and hence dust, and the chemistry of NaCl.}

\subsubsection{Pulsations}\label{sec:pulsations}

We tested whether the formation of the clumps could be partly driven by pulsations by estimating the time taken for each clump to reach its current location from the surface of the star. Even though the clump travel time from the stellar surface to a few stellar radii is uncertain, we are only considering the differences between clump travel times, so this region is cancelled out. We find the differences in travel times between clumps A, B and C to be close to \edits{integer multiples of the pulsation period}, within the uncertainties, which could suggest they were formed during similar pulsation phases. This does not hold for clumps D and E, however. Also, the spacing is uneven, such that we do not see a clump forming from each pulsation, making this result inconclusive. \edits{If clump E and the negative-$y$ protrusion of clump C are presently being shocked and only formed recently, that would suggest that a clump could be formed over several pulsations, since we estimate that the centre of clump C is 41 AU from the star (and the nearest edge, excluding the aforementioned protrusions, is only 3 AU from the star). Overall, we estimate that the difference in expansion time between the centres of clump C and clump D (the farthest clump) is $\sim20$--35~years, depending on the choice of expansion velocity.}
\edits{This is a remarkably short time period if we consider that the approximate amount of time it would take for the envelope to expand out to 1\arcsec\ \citep[the approximate radius of the HCN emission,][]{IKTau} is $\sim70$~years and the time taken to expand to the radius of the dust observed with PACS \citep[85\arcsec,][]{Cox2012} is around 6000 years. We don't see any NaCl beyond 0.6\arcsec{} or 150~AU from the star (Fig.~\ref{fig:areas} and \ref{fig:f(r)}), from which we can deduce that beyond this radius either the excitation conditions are not favourable for NaCl to be detected or the NaCl is removed from the gaseous phase, either because the molecule is destroyed or because it is condensed onto dust grains.}

\subsubsection{Continuum and dust}\label{sec:continuum}

\begin{figure*}
    \centering
    \includegraphics[width = 0.49\linewidth, draft = false]{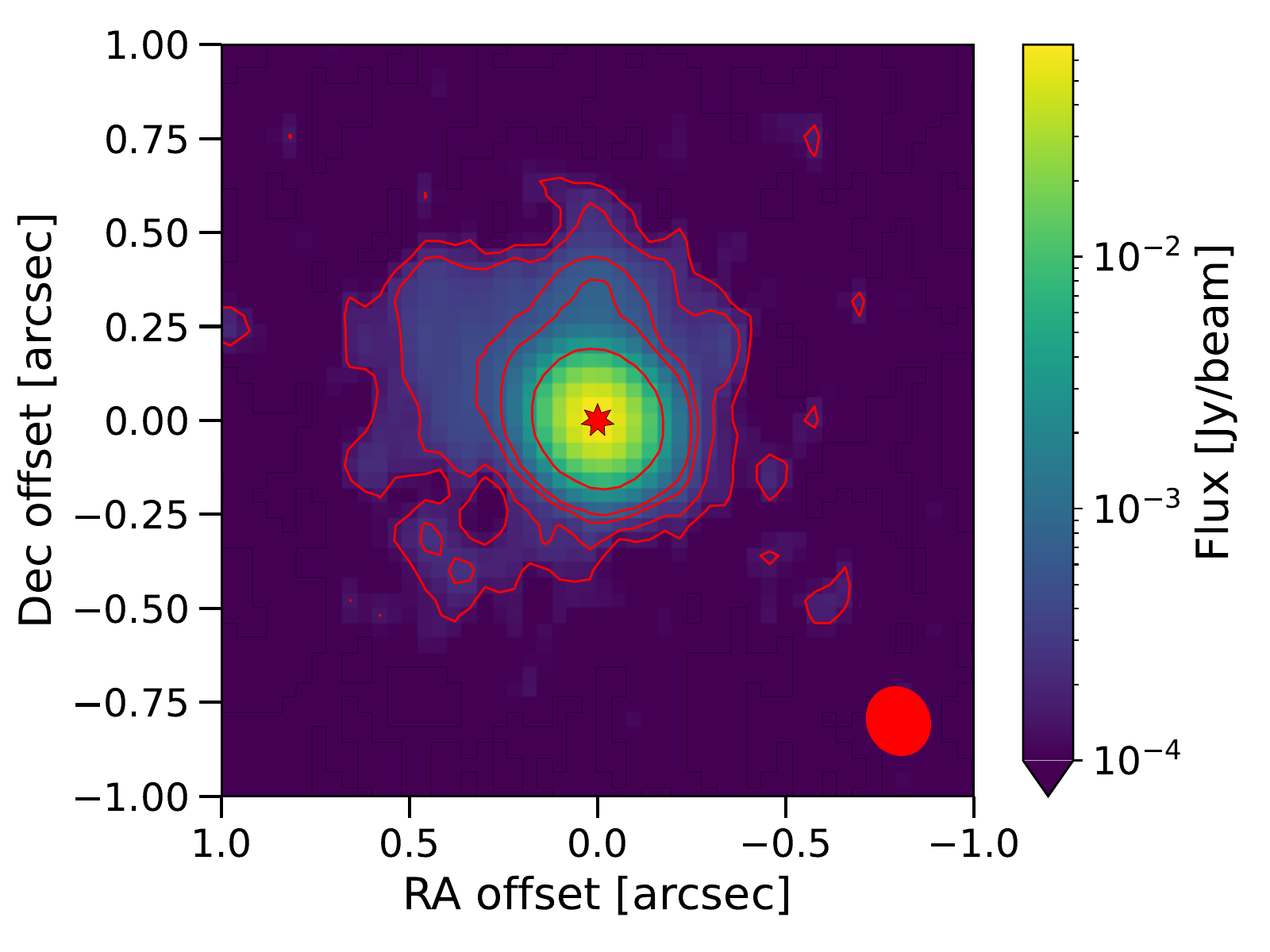}
    \includegraphics[width = 0.49\linewidth, draft = false]{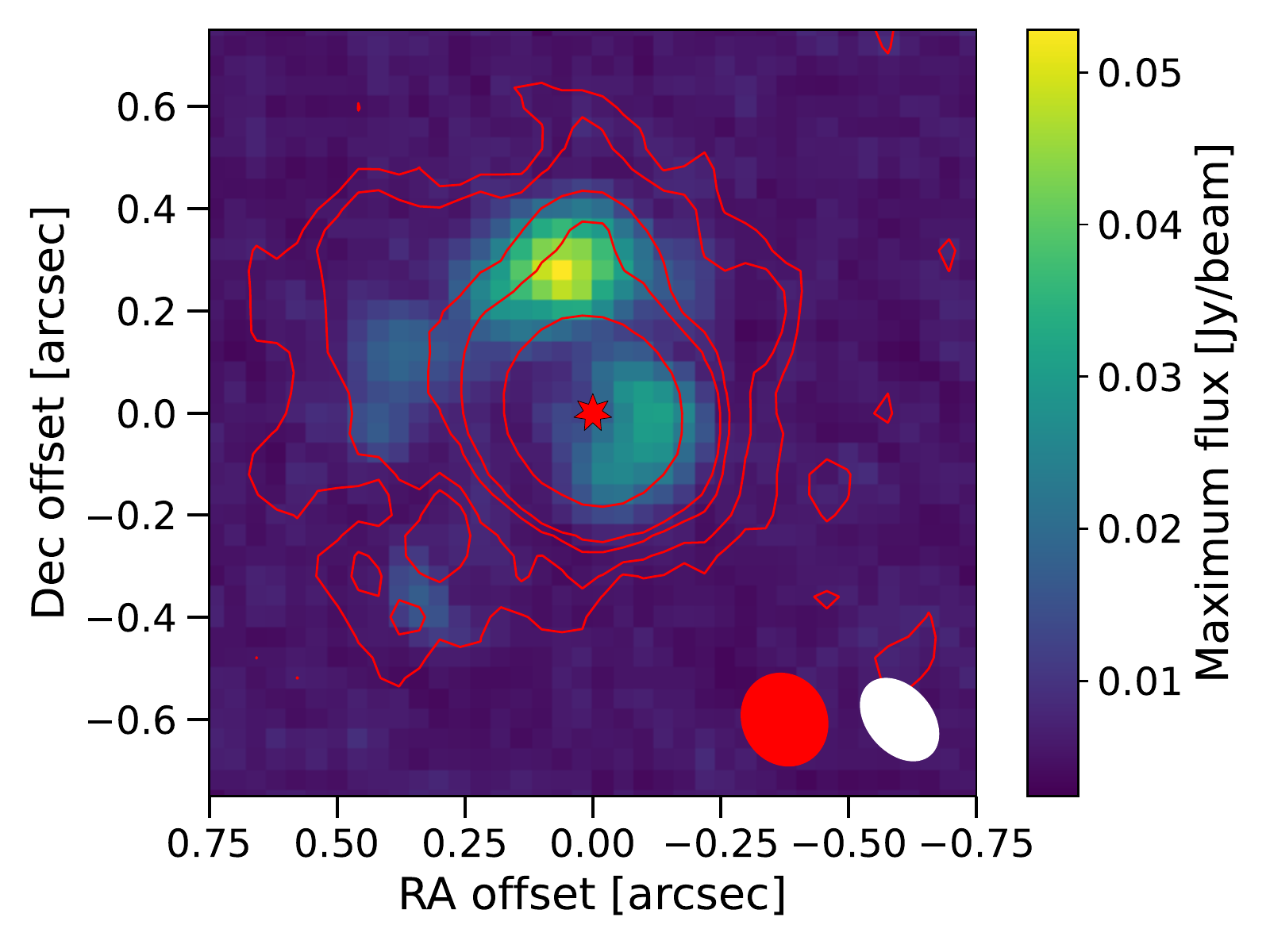}
    \caption{\edits{On the left we plot the continuum flux for IK Tau as observed by ALMA plotted on a log scale and with contours drawn at levels of 3, 5, 10, 15, and 100$\sigma$.  On the right we plot the maximum intensity plot of \transone, overlaid with the same red continuum contours. The white ellipse in the bottom right corner indicates the size of the synthetic beam for the NaCl emission. The red star indicates the location of the continuum peak and the red ellipse in the bottom right corner indicates the size of the synthetic beam. North is up and east is left.}
    }
    \label{fig:cont}
\end{figure*}

\edits{In Fig.~\ref{fig:cont} we plot the continuum emission of IK~Tau as observed with ALMA \citep[for details of the continuum data reduction, see][]{IKTau} Although the emission close to the star is almost circular (as shown by the $100\sigma$ contour), the contour at the $15\sigma$ level and the lower level contours show significant asymmetry. In the right panel of Fig.~\ref{fig:cont} we plot the same continuum contours over the maximum intensity plot of \transone. Hence, it becomes clear that the asymmetric continuum contours partly follow the spatial locations of the NaCl clumps. The $15\sigma$ continuum contour protrudes in the direction of clump A, the $10\sigma$ contour approximately encompasses clump A and protrudes in the direction of clump B, the $5\sigma$ contour encompasses clump B and the $3\sigma$ contour protrudes past clump D. There are also two small $5\sigma$ regions aligned with the clump D protrusion but located further from the star. This indicates that clump D is real but suggests that the excitation conditions for NaCl are such that clump D is too cool or distant from the star for \transtwo\ to be detected. We note also that the centre of clump D corresponds to a kinetic temperature of $\sim200$~K (Fig.~\ref{fig:nT}), which is already slightly lower than the $J=26$ level energy of 220~K. It is possible that lower-$J$ NaCl lines might result in a clearer detection of clump D and possibly even other cooler clumps that are not visible in either of the transitions covered by the present observations.}

\subsubsection{Chemical considerations}\label{sec:chem}

\editstwo{NaCl forms through the reaction of HCl with free sodium (\ce{Na + HCl \to NaCl + H}), which has been measured to be fast and with a high energy barrier \citep[$\sim 4100$~K,][]{Husain1986,Cherchneff2012}, meaning that it will form more readily at high temperatures.} In the absence of external forces \editstwo{on a spherically symmetric AGB wind}, NaCl is expected to form uniformly 
\editstwo{and at relatively low abundances ($\lesssim 10^{-8}$ relative to H$_2$)} 
in the stellar atmosphere \editstwo{and inner wind} at chemical equilibrium \citep[][and see further discussion in Sect.~\ref{sec:dis-f}]{Cherchneff2012,Agundez2020}, but such \editstwo{a uniform} scenario does not explain our observations. \editstwo{The chemical model of \cite{Gobrecht}, which includes shock-induced chemistry as a result of AGB pulsations, does not result in significantly higher abundances of NaCl (less than an order of magnitude increase) than the equilibrium chemical model of \cite{Agundez2020}.}

\editstwo{Recent studies have shown that dust and gas does not form isotropically in AGB stellar atmospheres, with large convective cells likely playing a role in gas and dust anisotropies close to the star \citep[][but see also \cite{Ohnaka2016,Takigawa2017,Khouri2018}]{Velilla-Prieto2023}. In principle, the anisotropic formation of NaCl could be explained by some dependence on convective cells, however, it is unclear why this would result in NaCl forming in only one region at a time, as we see in our data, or why there would be relatively long gaps between the enhancements in NaCl seen as clumps, as can be deduced from the distances given in Table \ref{tab:clump_dist} and briefly discussed in Sect.~\ref{sec:3D-struct} and \ref{sec:pulsations}. Although the spiral structure discussed in Sect.~\ref{sec:3D-struct} is only tentative, the emission is largely grouped in a plane in the CSE. Randomised ejections could form such a structure, but so could} something moving in a roughly circular trajectory around IK~Tau.
%
%

\editstwo{\cite{VY-CMa-1} proposed that the clumpy NaCl distribution towards VY CMa could be a result of sputtering off dust, a process that could be driven by shocks.}
\editstwo{For IK~Tau, the scenario of NaCl being formed on a single, rotating, location on the surface of the AGB star seems unlikely,}
so we suggest that the formation of NaCl \editstwo{could} be assisted by a body orbiting the star, which would also be a source of shocked gas if it were moving faster than the local sound speed, \editstwo{$\upsilon_{\text{sound}} \approx 3$ \kms}. \editstwo{Such a supersonic velocity} is likely since the orbital velocity for a planet or brown dwarf at 50~AU from IK~Tau would be 4--5 \kms{} \citep[based on the mass estimate for IK~Tau of][]{Danilovich2017}, higher than the sound speed in the innermost regions of the wind. For a circular orbit at 5~AU, this goes up to 14--16 \kms, well above the local sound speed. The shock conditions (i.e. higher temperatures and densities) generated by the orbit \editstwo{could} then drive the NaCl formation, \editstwo{either through the bimolecular chemical reaction of Na and HCl or through sputtering off dust, as suggested by \cite{VY-CMa-1}, which would also explain why the NaCl clumps correlate well with the dust seen in the ALMA continuum (Fig.~\ref{fig:cont})}. Peaks in the AGB stellar pulsation might also contribute to heightened shock conditions and hence increased NaCl formation (see Sect.~\ref{sec:pulsations}).
\editstwo{The atmospheric models of \cite{Bowen1988} predict that shocks from AGB pulsations will cause increases of a factor of a few in temperature and density in the inner wind, whereas the binary hydrodynamic models of \cite{Maes2021} and \cite{Malfait2021} predict increases up to an order of magnitude in density and temperature, depending on the model parameters. The hydrodynamic models of \cite{Aydi2022}, which consider both stellar pulsations and substellar companions in the inner wind, predict enhanced shock conditions when the shocked gas in the wake of a closely orbiting companion (orbital periods on the order of $\sim1000$~days) interacts with the shock from the pulsations of the AGB star. Such shocks also result in a shock velocity higher than the velocity of the ambient gas, which could explain the high velocities observed for clump E and the protrusion from clump C in the negative $y$-direction (see further discussion in Sect.~\ref{sec:3D-struct}), whereas (portions of) clumps that formed less recently have decelerated and cooled as they moved outwards in the circumstellar envelopes.
The models discussed here suggest that both stellar pulsations and shocks from orbiting companions could contribute to conditions which facilitate the production of NaCl.}
Collisions between the AGB wind and \editstwo{the shock generated by the supersonic orbit could also contribute to enhanced dust formation, potentially further enhanced during pulsations,} \citep[i.e. similar to, but less intense than, the process described by][for Wolf-Rayet stars]{Usov1991}, \editstwo{which could also explain} the continuum emission and why it correlates with the NaCl emission (see Fig.~\ref{fig:cont} and Sect.~\ref{sec:continuum}).

\edits{If the orbiting body is a star, it could also contribute to the heating of the gas through its radiation, \editstwo{further enabling NaCl formation by overcoming the formation energy barrier}, \editstwo{but such a star should not} be hot enough to destroy other molecules, since this is not seen in the observations. So a stellar companion to IK~Tau might be, for example, a cooler main sequence star, but not a white dwarf \citep[see a more detailed discussion of the contribution of stellar companions to the circumstellar chemistry of AGB stars in][]{Van-de-Sande2022}.
}

\editstwo{A less likely scenario is}
a rocky planetary body in a close orbit around the AGB star that is in the process of being destroyed. However, in such a case we would expect to see water and other metal-bearing species, \editstwo{composed of elements common to rocky planets}, in clumps similar to NaCl, which we do not \citep[e.g.][covering AlOH, AlCl, AlO, H$_2$O, and TiO, and note that no spatially resolved observations of cool water lines presently exist, but the observed hot water lines exhibit emission centred on the star, not in clumps]{Decin2017,IKTau,Danilovich2020a}. 
\editstwo{So we conclude that}, an orbiting body (planetary or substellar) \editstwo{would mainly} contribute to the aforementioned shock conditions in the wind, which then drive the formation of NaCl.


\subsection{NaCl abundance} \label{sec:dis-f}
The abundances and number densities we obtain for NaCl (see Table \ref{tab:f2}) depend on the pwv we used: for all clumps other than clump D, the obtained abundance is smaller at pwv = 1.7 mm than at pwv = 0.2 mm. Since a larger pwv value means that the simulated noise has a larger amplitude, the resulting lower abundance can be thought of as a correction for the higher noise. The clump abundances for the different pwv values all agree within the uncertainties, except for clump D. 
Fig. \ref{fig:synspec2} shows that the synthetic \transone\ spectra of clump D do not feature the peak that is visible in the observed spectrum, meaning that this clump isn't visible in the output. This is true whether or not the CASA ALMA simulator is applied to the synthetic channel maps (see Figs. \ref{fig:synmaps1} and \ref{fig:synmaps2}). This discrepancy can be explained by the fact that clump D was only detected in a single channel map in only one of the two observed transitions, meaning that there is only one data point in frequency or velocity space that isn't noise.

\begin{figure*}
    \centering
    \includegraphics[width = \linewidth, draft = false]{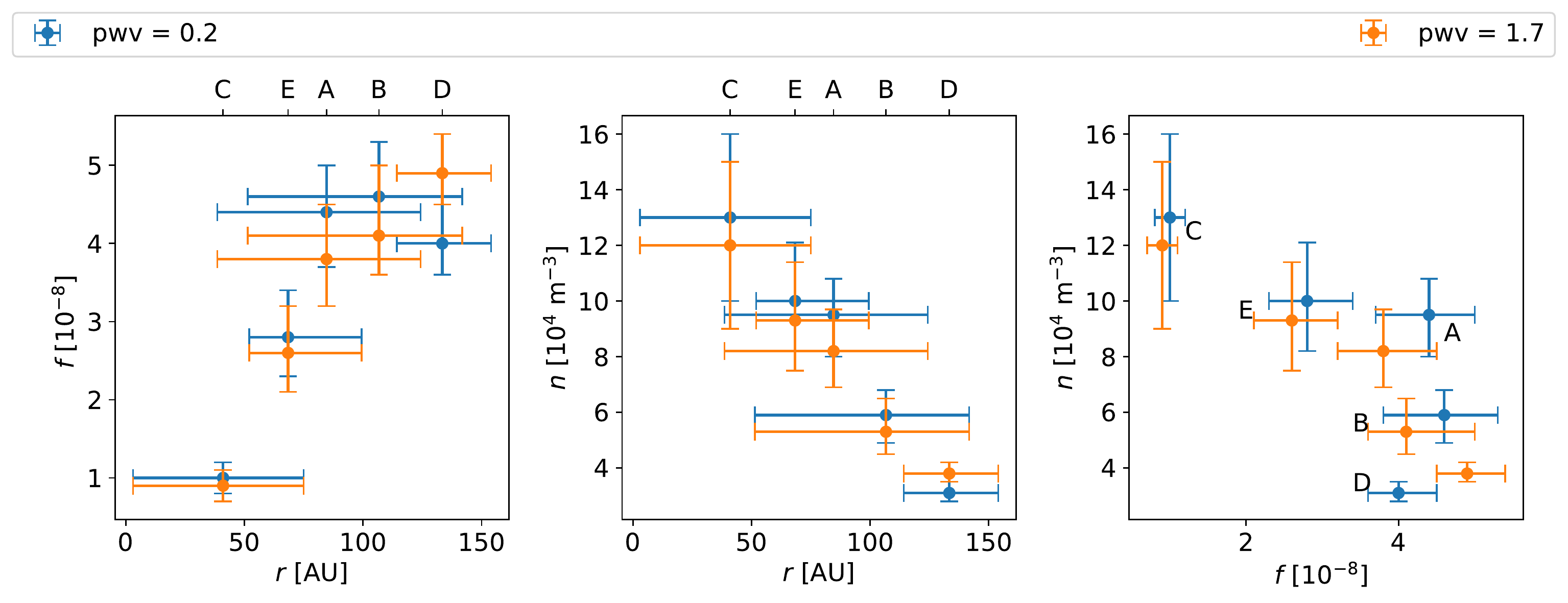}
    \caption{The relative NaCl abundances (left) and NaCl number densities (middle) of each of the clumps as a function of the distance of said clumps to the star. The right figure shows the NaCl number densities as a function of the relative NaCl abundances. The error bars on $f$ and $n$ represent the 50\% confidence interval, while the error bars on $r$ show the closest and farthest points of each clump.}
    \label{fig:f(r)}
\end{figure*}

In Fig. \ref{fig:f(r)} we plot the fractional abundances and number densities of each clump against the distance of each clump to the star, and also the fractional abundances against the number densities. The error bars on the radial distance $r$ indicate the nearest and farthest point of each clump to the star. This shows that while the clumps farther from the star have a higher fractional NaCl abundance (with the exception of clump D at pwv = 0.2 mm), the absolute NaCl number density does decrease with distance. This could mean that the NaCl number density decreases more slowly than the H$_2$ number density. It could also be an indication that NaCl formation has not run to completion in the clumps closer to the star. 

Comparing our abundances with the literature values in Table \ref{tab:NaCl_lit}, we can see that none of the clump abundances agree with the observation-based literature values that were found by \citet{VY-CMa-2} ($f\sim4\ee{-9}$) and \citet{Velilla-Prieto} ($f\sim3.1\ee{-7}$), which are respectively smaller and larger than our values ($f \sim 9\ee{-9} - 5\ee{-8}$). \citet{VY-CMa-2} used single dish data from the IRAM SMT and 12 m telescopes, while \citet{Velilla-Prieto} used single-dish data from the IRAM 30m telescope, and they both assumed a spherical distribution of NaCl with a source size of $0.3\arcsec$. In both datasets, the NaCl distribution isn't spatially resolved, and the assumed source size is too small, covering only a part of clump C rather than the entire emitting region. \citet{Velilla-Prieto} predicts that an underestimation of the emission region in their calculations would lead to an overestimation of their abundance, which we can confirm from our results.

\citet{Gobrecht} modelled non-equilibrium chemistry for IK~Tau in shocked gas layers close to the star. They found a range of NaCl fractional abundances from $f \sim 3.7\times10^{-10} - 2\times10^{-8}$ in the range $1R_\star$ -- $9R_\star$. These values only agree with our abundance for clump C, while we find the other clumps to have larger abundances. This is consistent with clump C lying partly in the region modelled by \citet{Gobrecht} ($r \in [1, 6]\Rstar$), \edits{including the shock where part of clump C is forming,}  while the other clumps we observed lie further from the star. \edits{The difference between our result and the \cite{Gobrecht} model can also be explained by a companion orbiting IK~Tau, as suggested in Sect~\ref{sec:chem}, because the shock generated by the companion is not included in the chemical model.}
\citet{Agundez2020} calculated the thermochemical equilibrium abundances of molecules in the atmospheres of M-type AGB stars, between radial distances of 1 \Rstar\ to 10 \Rstar. At the outer part of their model, ($r = 10 \Rstar$), they find an NaCl abundance of $f \sim 1\ee{-9}$ relative to H$_2$. A part of clump C lies within 10 \Rstar, but the fractional abundance we find for clump C is almost one order of magnitude larger. This is further evidence that equilibrium conditions are not responsible for the formation of NaCl \edits{and underscores the importance of shock chemistry}.

\section{Conclusions}\label{sec:conclusions}
In this paper, we studied the clumpy distribution of NaCl in the CSE of the AGB star IK Tau. First, a three-dimensional model of the spatial distribution of NaCl around the star was obtained by deprojecting the channel maps of the \transone\ transition observed with the ALMA telescope. To do this, we derived a new value for the maximum expansion velocity $\vinf = 28.4 \pm 1.7$ \kms, using the SiS $J=19\rightarrow 18$, $v=0$ emission line in the spectrum of IK~Tau. 
The three-dimensional model indicates that each of the clumps has a different shape and size, and is located at a different distance from the star. 
The overall distribution of these clumps suggests a tentative spiral shape, and the relative size of these clumps and their distances from the central star suggest that the farthest clumps were formed earliest and are now smaller either because of dust condensation or because they expand while moving outwards, making their outer regions too diluted to be detected. \edits{We speculate that the formation of NaCl is driven by shocks originating from stellar pulsations and/or an orbiting body, explaining the clumpy spiral distribution.}

We derived the NaCl abundances of each of the clumps using the three-dimensional radiative transfer modelling code \magritte, with our deprojected model as input. This modelling was performed under the assumption of LTE to keep the computational times reasonable. The clump abundances lie in the range $9\ee{-9}$ -- $5\ee{-8}$ relative to H$_2$. The abundance of the clump closest to the star falls within the range predicted by shock-induced chemistry models. The higher relative abundances of the clumps further from the star indicate that NaCl formation may continue at larger distances. 


\begin{acknowledgements}
\editstwo{The authors grateful to the anonymous referee for providing many constructive comments. We also thank Daniel Price for fruitful discussions regarding shocks.} This paper makes use of the following ALMA data: ADS/JAO.ALMA2013.1.00166.S. ALMA is a partnership of ESO (representing its member states), NSF (USA) and NINS (Japan), together with NRC (Canada) and NSC and ASIAA (Taiwan), in cooperation with the Republic of Chile. The Joint ALMA Observatory is operated by ESO, AUI/NRAO and NAOJ. Credit CASA: International consortium of scientists based at the National Radio Astronomical Observatory (NRAO), the European Southern Observatory (ESO), the National Astronomical Observatory of Japan (NAOJ), the CSIRO Australia Telescope National Facility (CSIRO/ATNF), and the Netherlands Institute for Radio Astronomy (ASTRON) under the guidance of NRAO.

TD and FDC acknowledge support from the Research Foundation Flanders (FWO) through grants 12N9920N, and 1253223N, respectively. TD is supported in part by the Australian Research Council through a Discovery Early Career Researcher Award (DE230100183). LD thanks the support of the Fund of Scientific Research via the Senior Research Project G099720N.
\end{acknowledgements}

%
%

\bibliographystyle{aa} 
\bibliography{bibliography} 

\begin{appendix}
\section{Additional plots of NaCl}\label{sec:A}

\subsection{Spectral lines}

\begin{figure*}[t]
    \centering
    \includegraphics[width = \linewidth]{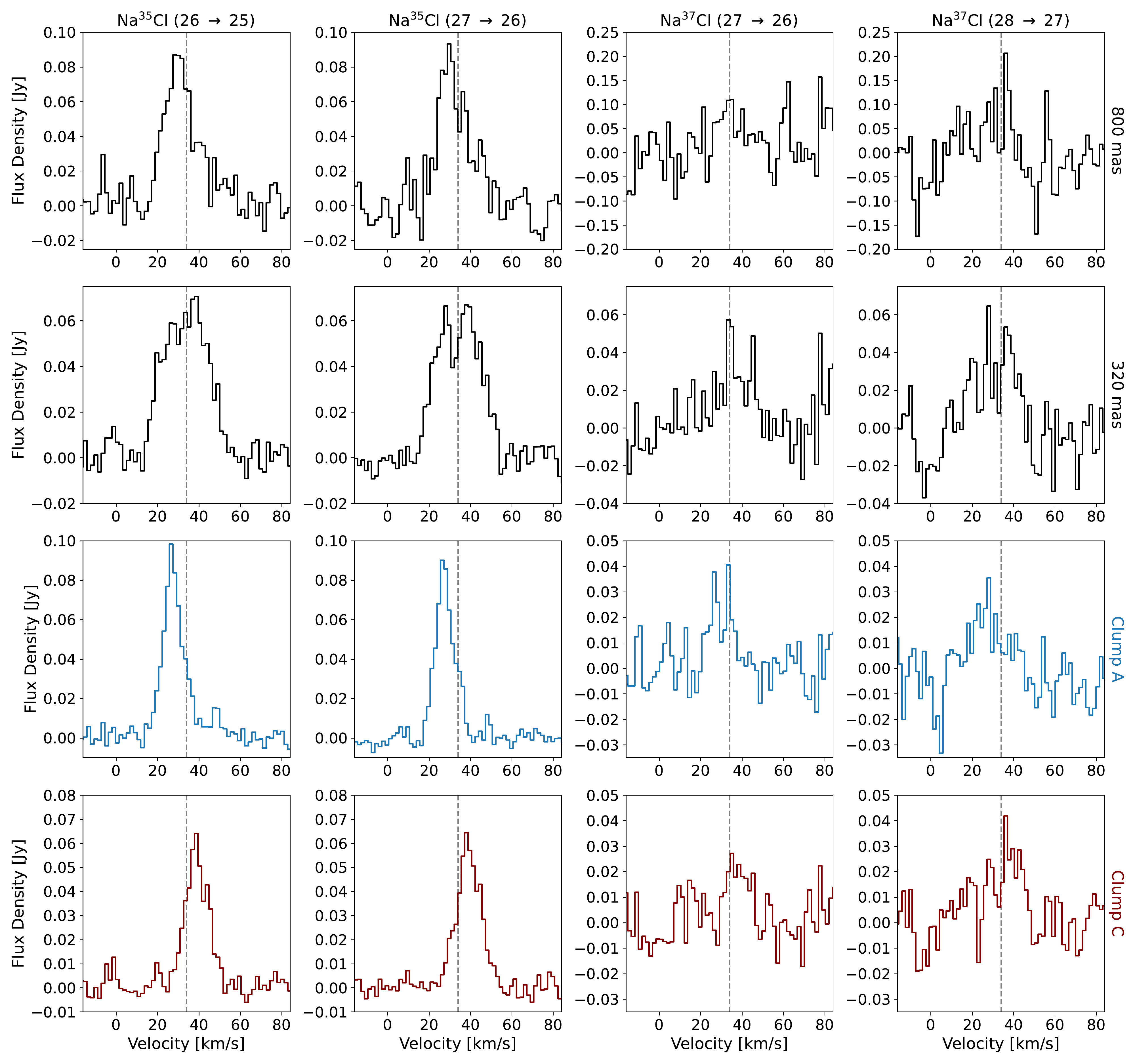}
    \caption{\edits{Plots of the four spectral lines of NaCl covered by ALMA. The transitions are given at the top of each column and the extraction apertures at the right of each row. The spectra in the top two (black) rows were extracted for circular apertures (with radii of 800 and 320~mas) centred on the star, while the bottom two rows show spectra extracted for irregular apertures encompassing clumps A and C respectively. The dashed grey vertical lines indicate $\upsilon_\mathrm{LSR}=34$~\kms.}}
    \label{fig:allspec} 
\end{figure*}

\edits{In Fig.~\ref{fig:allspec}, we plot all four lines of NaCl covered by the ALMA spectral scan of IK~Tau \cite{IKTau} using four different extraction apertures: circular apertures with radii of 800~mas and 320~mas centred on the star, and two irregular elliptical apertures, chosen in CASA to capture the majority of the emission from clumps A and C, the two brightest clumps. Further discussion can be found in Sect.~\ref{sec:observations}.}

\subsection{Observed channel maps}
\begin{figure*}
    \centering
    \includegraphics[width = \textwidth]{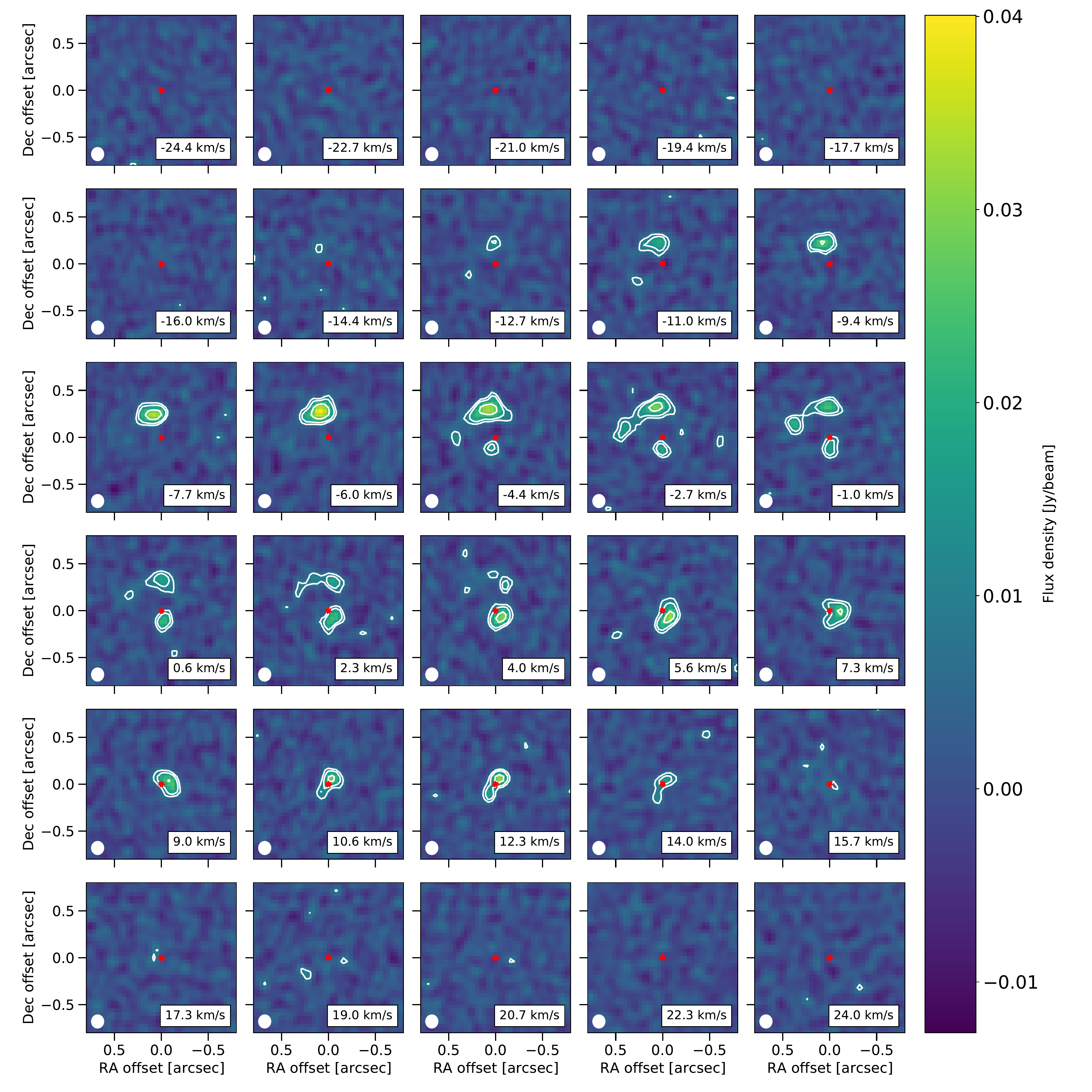}
    \caption{Channel maps of the \transtwo\ line. The red star indicates IK Tau's position, and the white ellipse shows the beam size. The white contours are at $3\sigrms$, $5\sigrms$, and $10\sigrms$ ($\sigrms = 2.6$ mJy). The $\vlsr = 34$ \kms{} has been subtracted from the  velocities.}
    \label{fig:channelmaps2} 
\end{figure*}

Fig. \ref{fig:channelmaps2} shows the observed channel maps of the \mbox{\transtwo} transition. This figure is discussed in Sect. \ref{sec:observations}.
\subsection{Synthetic channel maps}
\begin{figure*}[h!]
    \centering
    \includegraphics[width = \linewidth, draft = false]{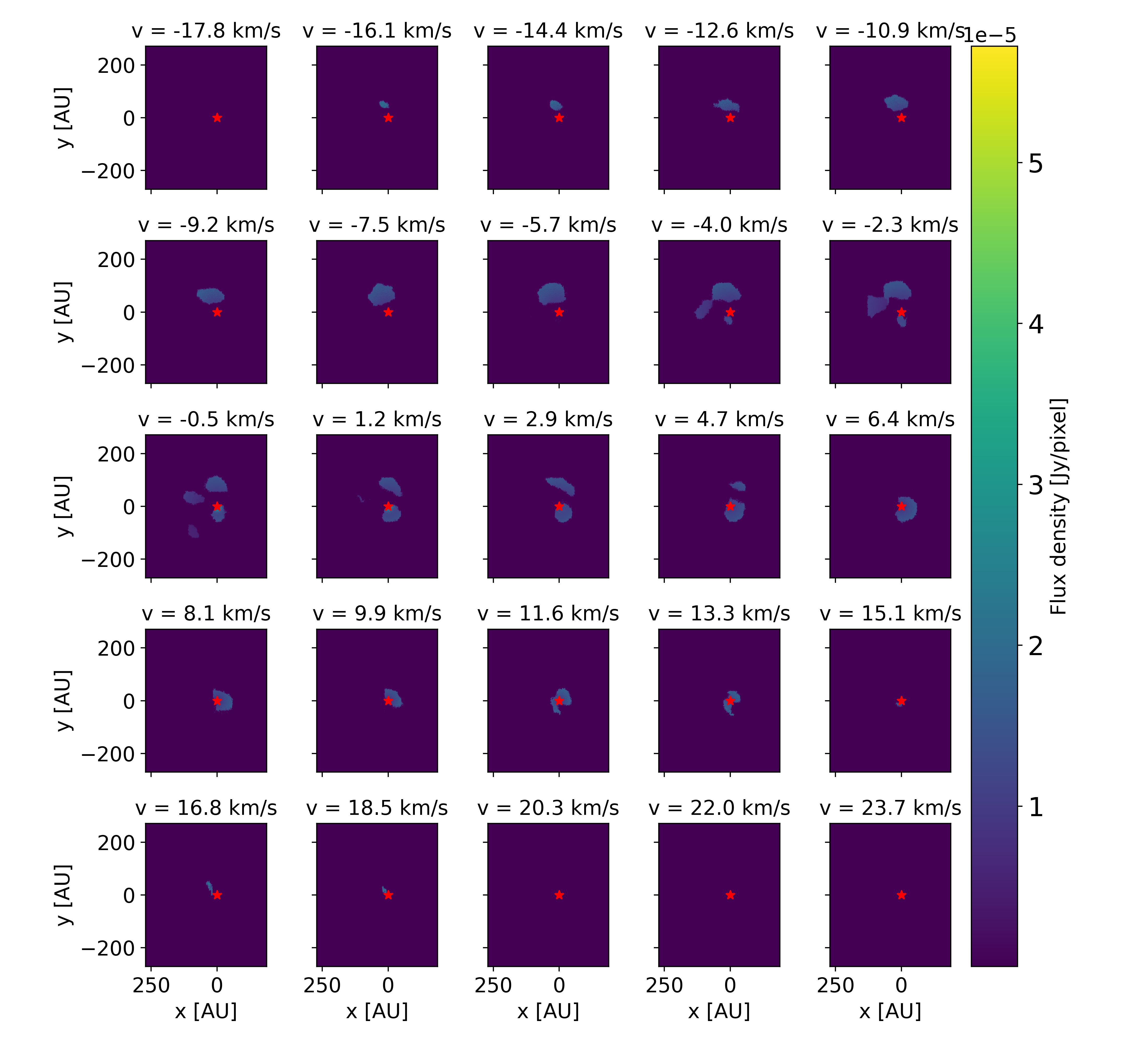}
    \caption{Synthetic \magritte\ channel maps of the \transone\ line, after the observed velocity bins are recreated by combining multiple maps together. The red star indicates IK Tau's position.}
    \label{fig:synmaps1} 
\end{figure*}

\begin{figure*}[h!]
    \centering
    \includegraphics[width = \linewidth, draft = false]{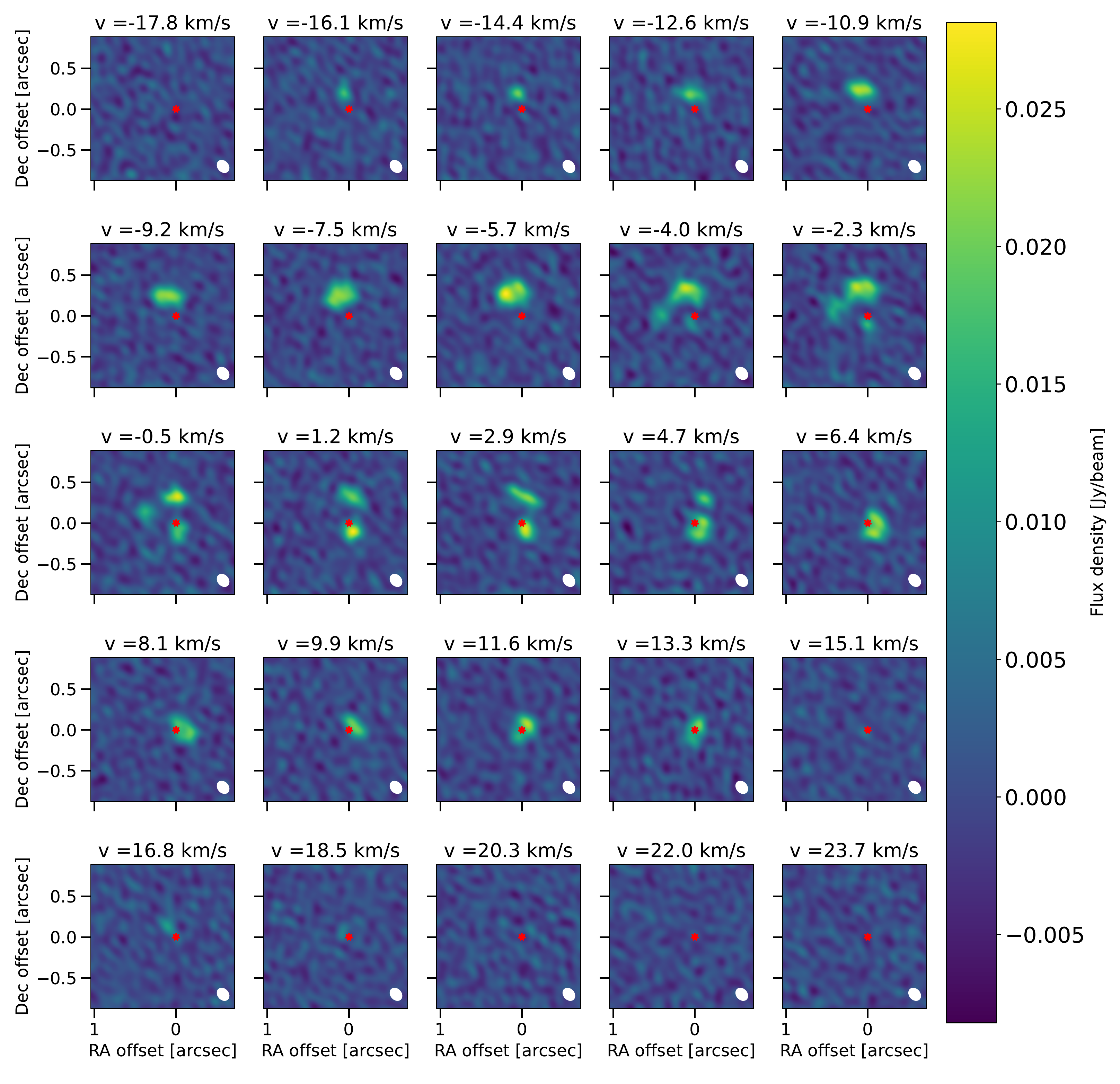}
    \caption{Synthetic \magritte\ channel maps of the \transone\ line from Fig. \ref{fig:synmaps1}, after the CASA simulator has been applied to them, using pwv = 0.2 mm. The red star indicates IK Tau's position, and thee white ellipse shows the (simulated) beam size.}
    \label{fig:synmaps2} 
\end{figure*}

Figs. \ref{fig:synmaps1} and \ref{fig:synmaps2} both show the synthetic channel maps of the same model, before and after the CASA simulator has been applied to them respectively. These figures are mentioned in Sects. \ref{sec:RTM} and \ref{sec:dis-f}.
\section{Spectral lines of CO and SiS}\label{sec:B}
\begin{figure*}[h!]
    \centering
    \includegraphics[width = \linewidth]{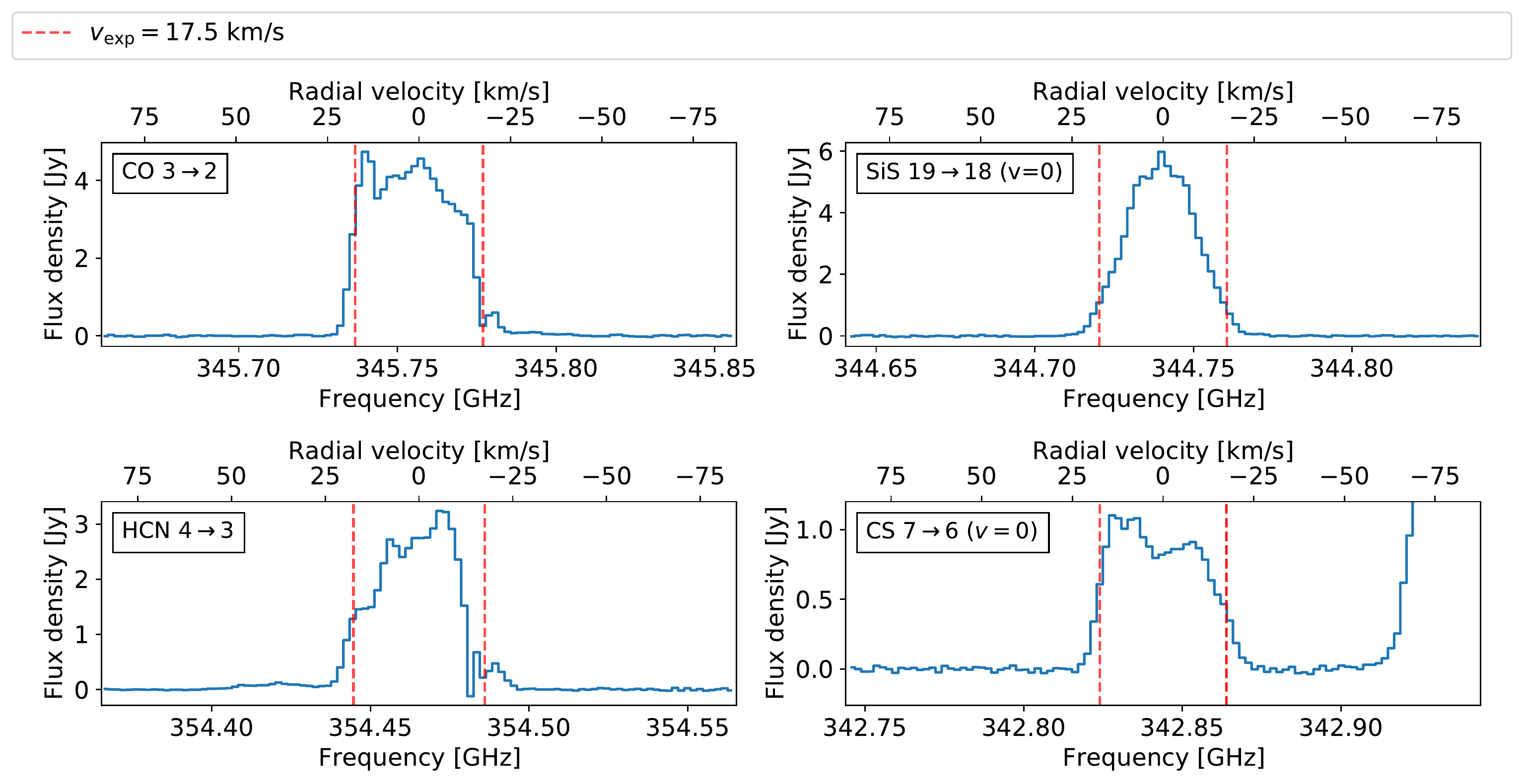}
    \caption{\edits{The spectral lines corresponding to the CO 3 $\rightarrow$ 2 (\textit{top left}), 
    SiS 19 $\rightarrow$ 18 ($v$=0) (\textit{top right}), 
    HCN  4 $\rightarrow$ 3 (\textit{bottom left}),
    CS 7 $\rightarrow$ 6 ($v$ = 0) (\textit{bottom right}) transitions. 
    The red dashed lines indicate the $\vinf = \pm 17.5$ \kms. The used aperture has a radius of 320 mas.}}
    \label{fig:otherspectra} 
\end{figure*}

Fig. \ref{fig:otherspectra} shows the observed spectral lines of the CO 3 $\rightarrow$ 2 and SiS 19 $\rightarrow$ 18 ($v$=0) transitions. They are used in Sect. \ref{sec:velo} to determine the maximum expansion velocity of IK Tau.

\edits{\section{Deprojection} \label{sec:appendix-deprojection}
\subsection{Derivation of the deprojection equation} \label{sec:derivation}
An equation for the $z$ coordinate can be derived from the fact that the observed channel map velocity is the component of the expansion velocity along our line of sight. 
If we assume a spherically symmetric stellar wind with a constant expansion velocity $\vinf$, we can then define $\theta$ as the angle between the channel map velocity $\vchannel$, from which the $\vlsr$ has been subtracted, and the expansion velocity, $\vinf$, such that $\vchannel = \vinf\cos(\theta)$. This is illustrated in Fig.~\ref{fig:deproj_diagram}. The expression can then be rewritten as
\begin{equation}\label{eq:tan(theta)1}
    \tan(\theta) = \frac{\sqrt{\vinf^2 - \vchannel^2}}{\vchannel}
\end{equation}
For any point on a particular channel map, this angle $\theta$ also connects the coordinate along our line of sight $z$ with the radial distance $r$ to said point, such that $z = r\cos(\theta)$, as shown in Fig.~\ref{fig:deproj_diagram}.
Using the impact parameter $p = \sqrt{x^2 + y^2} = r\sin(\theta)$, we can write the following expression:
\begin{equation}\label{eq:tan(theta)2}
    \frac{p}{z} = \tan(\theta)
\end{equation}
Combining equations \eqref{eq:tan(theta)1} and \eqref{eq:tan(theta)2}, yields us the expression for the coordinate along our line of sight:
\begin{equation} 
    z = \sqrt{x^2 + y^2}\frac{\vchannel}{\sqrt{\vinf^2-\vchannel^2}}
\end{equation} 
\begin{figure}
    \centering
    \includegraphics[width = \linewidth]{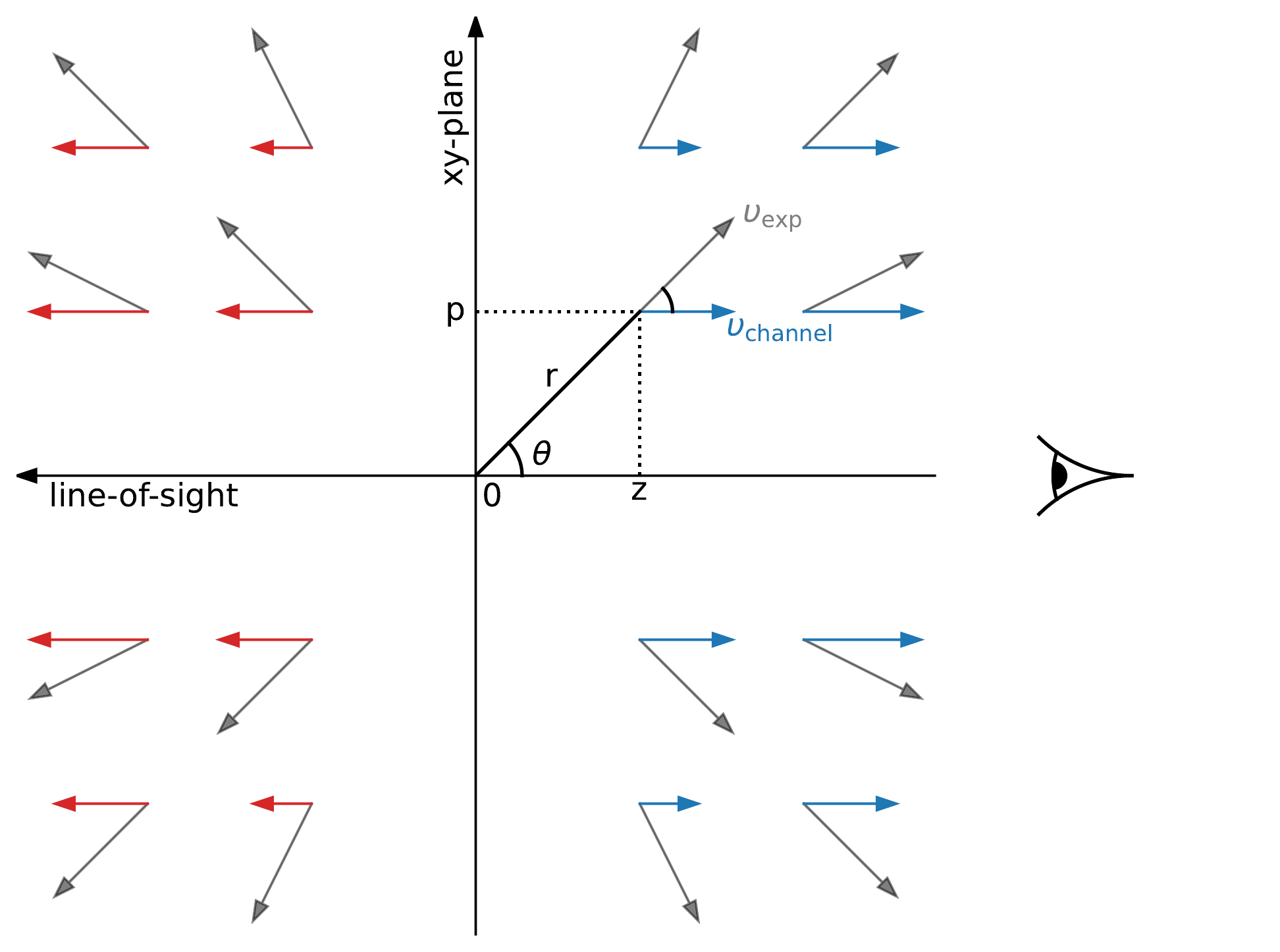}
    \caption{\edits{Diagram illustrating the principles of deprojection from radial velocity to spacial coordinates, using a constant radial expansion velocity $\vinf$. The observer is located on the right. The impact parameter is $p = \sqrt{x^2 + y^2}$.}}
    \label{fig:deproj_diagram} 
\end{figure}%
We also note that from the definitions of $\vchannel$ and $z$, that this equation can also be written as
\begin{equation}
    z = \frac{\vchannel}{\vinf}r\;.
\end{equation}
\subsection{Deprojection of planes of constant velocity along the line of sight}
Fig. \ref{fig:deproj_principle} shows the results of a general deprojection. In this figure, five planes of constant velocity along the line of sight were deprojected, that were evenly spaced within velocity space. Only the positive $x$ side of each plane is shown. On each plane, the points with a small impact parameter $p = \sqrt{x^2+y^2}$ are mapped to smaller $z$ values, and thus closer to the star, while points with a larger impact parameter $p$ are mapped to much greater $z$ values. The larger the velocity associated with the plane, the larger the spread in $z$ values will be. Since every channel map has a constant velocity everywhere, the deprojections of these planes are a good representation of how the contents of the channel maps are deprojected: on a single channel map, the points that lie close to the star will be deprojected to smaller $z$ values than the points that lie further from the map. This effect is enhanced for channel maps with velocities that are much greater than zero.
\begin{figure*}
    \centering
    \includegraphics[width = \linewidth]{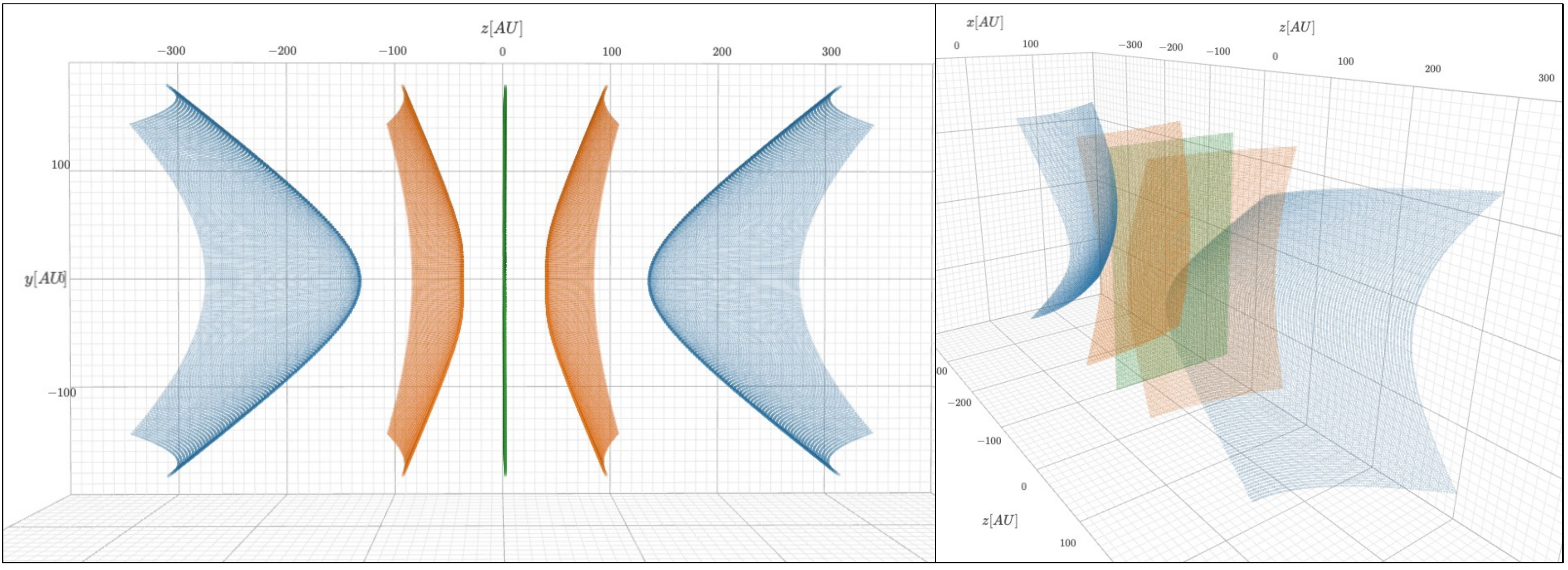}
    \caption{\edits{Two three-dimensional images of five deprojected planes, where every point within a plane has the same velocity. Planes with the same colour have the same velocity with opposite sign. The planes are only plotted for positive $x$ values.}}
    \label{fig:deproj_principle} 
\end{figure*}%
}

\section{Excitation analysis}\label{sec:RTD}
\subsection{Calculation}
Rotational temperature diagrams or population diagrams can be used to deduce rotational temperatures and column densities \citep{Goldsmith_1999}. This method assumes that the emitting region is in LTE and is optically thin. We expect NaCl to be optically thin, because we expect relatively low abundances ($\lesssim 10^{-7}$). According to the \magritte{} modelling, the maximum optical depth is $\tau = 6\ee{-5}$. The assumption of LTE might not hold for the clumps, due to the low density of the CSE ($\sim 10^{11}-10^{12}$~cm$^{-3}$) and the high Einstein $A$-values. 
The rotational temperature $T_{\text{rot}}$ is the excitation temperature of the purely rotational transitions.  

In LTE, the column density of each level is proportional to its population, and can be expressed as:
\begin{equation}
    N_u = \frac{N}{Z}g_ue^{-E_u/k\Trot}
\end{equation}
Here, $N_u$ is the column density of the upper level $u$ in cm\textsuperscript{-2}, $N$ is the total column density of this particular species in cm\textsuperscript{-2}, $Z = \sum_{i} N_i$ is the partition function, $g_u$ is the statistical weight of the level $u$, $E_u$ is the energy of the level $u$ in eV, $k$ is the Boltzmann constant in eV K$^{-1}$ and $\Trot$ is the excitation temperature in K. 
 
This equation can then be rewritten as:
\begin{equation}\label{eq:population_diagram}
    \ln\frac{N_u}{g_u} = \ln\frac{N}{Z} - \frac{1}{\Trot}\frac{E_u}{k}
\end{equation}
The rotational temperature diagram or population diagram is then a scatter plot with $(E_u/k)$ on the x-axis and $\ln(N_u/g_u)$ on the y-axis. Under the assumptions of LTE and optically thin lines, the datapoints should form more or less a straight line. The rotational temperature $\Trot$ can then be found from the slope, and the column density $N$ can be found from the intercept (as long as the partition function $Z$ is known).
We use this method to obtain the rotational temperatures of the NaCl clumps.

To calculate $N_u$ for each transition, the method of \cite{rot_temp} was used. Here, $N_u$ is defined as:
\begin{equation}
    N_u = \left(\frac{4\pi D^2}{\pi r_e^2}\right)\left(\frac{W}{A_{ul}h\nu}\right)
\end{equation}
Here, $D$ is the distance to the star, $r_e$ is the radius of the emitting region, $W$ is the line flux in units (W/m$^2$), $A_{ul}$ is the Einstein coefficient of spontaneous emission for a transition linking levels $u$ and $l$, $h$ is Planck's constant and $\nu$ is the frequency of the transition. Fig. \ref{fig:areas} shows the locations of the emission regions for clumps A, B and C, and for a larger region encompassing all the emission. We exclude clumps D and E from the rotational diagram analysis because they are not seen above the noise in the \transtwo\ channel maps.
The emission regions are ellipses with major and minor axes given in table \ref{tab:Trot}.
$W$ is found by integrating over the spectral line extracted for each emission region. An error of $7\%$ on the flux density \citep{IKTau} and an error of 0.02\arcsec\ on the radii of the emission regions are taken into account to find the uncertainties on the rotational temperature.
\begin{figure}
    \centering
    \includegraphics[width = \linewidth]{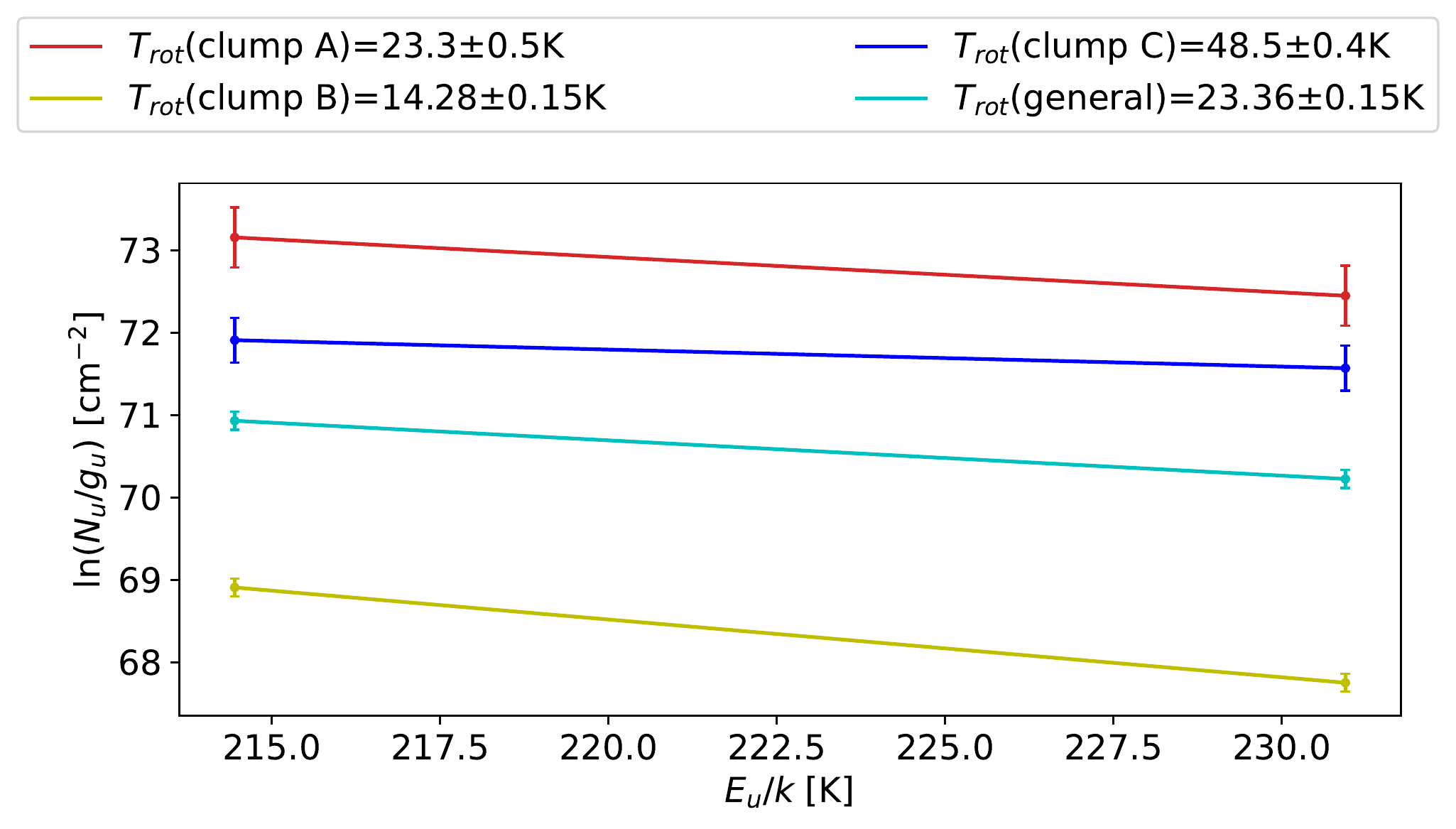}
    \caption{The rotational temperature diagram for four different emitting regions of NaCl around IK Tau.}
    \label{fig:Trot} 
\end{figure}%
Fig. \ref{fig:Trot} shows the rotational temperature diagram calculated using the regions shown in Fig. \ref{fig:areas}. We find rotational temperatures rainging from \mbox{14 -- 49 K}, as listed in table \ref{tab:Trot}.
The uncertainties given in this table are underestimates, however, because there are only two datapoints, and thus the slopes were calculated directly rather than fitted. If more NaCl lines were observed, more accurate rotational temperatures could be derived.
\begin{table}[]
    \centering
    \caption{{Extraction aperture sizes and rotational temperatures for each clump}.}
    \begin{tabular}{c|ccc}
    \hline\hline
         Region &Major axis [\arcsec] & Minor axis [\arcsec]& $T_{\text{rot}}$ [K] \\
         \hline
         Clump A &0.23&0.15& $23.3\pm0.5$ \\
         Clump B &0.16&0.12& $14.3\pm0.2$ \\
         Clump C &0.17&0.12& $48.5\pm0.4$ \\
         General &1.2\phantom{1}&1.2\phantom{1}& $23.4\pm0.2$ \\
         \hline
    \end{tabular}
    \label{tab:Trot}
\end{table}

\subsection{Discussion} 
Comparing the $T_{\text{rot}}$ of the three different clumps, it becomes clear that the $T_{\text{rot}}$ is larger for clumps nearer to the star than for those farther from the star. This makes sense, since the physical temperature is highest near the star. That the temperature of the entire emission region lies near the average of the three clumps makes sense as well, since it takes the entire region into account. Our results are very sensitive to noise, because they were derived from only two datapoints, because there were only two NaCl transitions in the observed range. It is thus necessary to observe more transitions before conclusive results can be obtained.

\citet{Velilla-Prieto} determined a rotational temperature of $T_{\text{rot}} = 67 \pm 7$ K for NaCl, using single-dish data from the IRAM 30m telescope, and assumed a circular NaCl emitting region with a diameter of $0.3\arcsec$. 
The disagreement between their and our results can be explained by two factors, the first being that, as mentioned before, in their observations, the NaCl distribution isn't spatially resolved and their assumed emission region only covers a part of clump C. A second factor is that we only had two datapoints available, while they used 13, but the first factor overshadows this benefit.
The diameter of their assumed emission region is 4 times smaller than the diameter of our general emission region, making its area too small by a factor of 16. They were thus unable to take most of the emission into account for their calculations.
Interestingly, the $T_{\text{rot}}$ we derived for the separate clumps lie in a similar range as the $T_{\text{rot}}$ found by \cite{Velilla-Prieto} for most other molecules in the CSE of IK Tau. They found that most molecules, other than NaCl and SiS, display a rotational temperature between 15 K and 40 K in the CSE of IK Tau. Whether this means that the $T_{\text{rot}}$ of NaCl should actually be closer to this range can only be determined with more spatially resolved observations of more NaCl lines.

Finally, the derived rotational temperatures are much lower than those of the kinetic temperature profile of IK Tau derived by \cite{velprofile}, at similar radial distances (this temperature profile is shown in Fig. \ref{fig:nT}). This is especially so for clump C, which lies very close to the star, yet has a rather cool rotational temperature of $T_{\text{rot}} = 48.5\pm0.4$ K. 
This large discrepancy indicates that the assumption of LTE does not hold, as in LTE, the rotational temperature is equal to the gas kinetic temperature \citep{rot_temp}. These results should be further verified by observing more spatially resolved transitions of NaCl.

\end{appendix}

\end{document}